\documentclass[11pt]{article}
\usepackage[utf8]{inputenc}
\usepackage[a4paper,margin=1in]{geometry}
\usepackage{setspace}
\usepackage[dvipsnames,table,xcdraw]{xcolor}
\usepackage{graphicx}
\usepackage[percent]{overpic}
\newcommand{\jrsscpanellabel}[1]{{\normalfont\normalsize\bfseries(#1)}}
\usepackage[colorlinks,citecolor=blue,urlcolor=blue,linkcolor=blue]{hyperref}
\usepackage{subcaption}
\usepackage{mathtools}
\usepackage{bm}
\usepackage{natbib}
\usepackage{amsmath}
\usepackage{amssymb}
\usepackage[labelfont=bf,font=small]{caption}
\usepackage{booktabs}
\usepackage{longtable}
\usepackage{array}
\usepackage{multirow}
\usepackage{float}
\usepackage{orcidlink}

\newcommand{\bs}{\boldsymbol}

\DeclareMathOperator{\Cov}{Cov}
\DeclareMathOperator{\diag}{diag}

\providecommand{\keywords}[1]{%
  \small\textbf{Keywords ---} #1
}

\title{
\textbf{Borrowing Space, Clustering Temporal Dynamics: Bayesian Regime Mapping and Spatial Prediction for European PM10}
}

\author{
  Luca Aiello\textsuperscript{1,a} and
  Lucia Clarotto\textsuperscript{2,b}
  \\ 
  \scriptsize \textsuperscript{1}Department of Biostatistics, University of California, Los Angeles, Los Angeles, CA, USA, \textsuperscript{a}\href{mailto:laiello@g.ucla.edu}{laiello@g.ucla.edu} \\
  \scriptsize \textsuperscript{2}UMR MIA Paris-Saclay, AgroParisTech, INRAE, Universit\'e Paris-Saclay, Palaiseau, France, \textsuperscript{b}\href{mailto:lucia.clarotto@agroparistech.fr}{lucia.clarotto@agroparistech.fr}
}

\date{}

\begin{document}

\maketitle

\begin{abstract}
Daily particulate matter measurements from European monitoring networks vary both through spatially shared fluctuations and through differences in local temporal dynamics. Characterising both features is important for identifying groups of monitoring locations with similar temporal behaviour and for predicting concentrations at unmonitored sites. We develop a Bayesian model that combines a continuous spatial common-shock process with a spatial product partition prior for clustering locations according to their local persistence and innovation variability. At unmonitored locations, the model extends these components through spatial interpolation and probabilistic regime allocation. Sparse spatial computations and online prediction summaries make inference feasible for 2,989 monitoring locations and 5,297 unmonitored land locations over 2024. Replicated simulations show that geographic information improves recovery of weakly separated, spatially organised regimes, while uncertainty in regime allocation need not prevent useful continuous prediction. The European analysis identifies stable clusters spanning national boundaries. Clusters with similar persistence nevertheless differ in local innovation variability and in how consistently their stations are grouped together across posterior draws. Comparisons with three restricted models clarify the complementary roles of spatially shared variation, heterogeneous local dynamics and geographic information in the partition.
\end{abstract}


\section{Introduction}
\label{sec:introduction}

European air-quality networks provide daily measurements at irregularly distributed monitoring locations. A concentration map describes where pollution is higher or lower, but does not reveal whether residual departures at different sites have similar persistence or local variability. Conversely, grouping sites by their temporal behaviour does not by itself provide a continuously varying field for prediction between monitoring locations. Nearby sites may share fluctuations while differing in local dynamics, and distant sites may exhibit similar dynamics without strong contemporaneous dependence. The applied problem is therefore to characterise these complementary structures without treating temporal-regime groups as homogeneous pollution regions. This motivates the joint regime-mapping and spatial-prediction framework developed here, building on dynamic and spatio-temporal environmental models \citep{gelfand2005spatial,sahu2007high,cressie2011statistics,cameletti2013spatio,pereira2026spde}.

Model-based clustering provides a natural way to characterise heterogeneous time-series dynamics \citep{fruhwirth2008model,nieto2014bayesian}. For spatially referenced series, related approaches include dynamic mixtures \citep{paci2018,cheam2017}, spatially constrained or partition-based clustering \citep{teixeira2019,zhong2026bayesian}, and spatially dependent Bayesian nonparametric constructions \citep{duan2007generalized,bucci2022,jung2024}. Our starting point is the parameter-based formulation of \citet{aiello2026bayesian}, which clusters sites through interpretable parameters governing their local temporal dynamics. Product partition models (PPMs) \citep{hartigan1990partition,quintana2003bayesian} and their spatial extension (sPPMs) \citep{page2016spatial} provide a flexible prior on an unknown partition while encouraging geographically coherent groups. However, on one hand, clustering local dynamics alone does not provide a continuously varying spatial process for prediction between monitoring sites. On the other hand, forcing all spatial dependence into the partition can blur the distinction between smooth shared variation and discrete regime differences.

We address this problem with the Decoupled Common-Shock spatial Product Partition Model (DCS-sPPM). The methodological contribution is a joint construction in which continuous spatial dependence and discrete heterogeneity in local temporal dynamics play distinct inferential roles. A common-shock process represents residual fluctuations shared across sites, while a spatial product partition prior groups locations through their persistence and innovation variance after accounting for that shared component. This distinction also governs prediction: shared variation is extended through spatial interpolation, whereas local dynamics are extended through probabilistic regime allocation. A nearest-neighbour Gaussian process (NNGP) representation \citep{datta2016hierarchical}, based on sparse conditional approximations \citep{vecchia1988estimation}, and online predictive summaries make the joint analysis feasible for a continental monitoring network. The contribution lies in this integrated inferential construction, rather than in any of these established ingredients individually.

The resulting posterior provides two complementary summaries at an unmonitored location: a continuously varying environmental process and a distribution over local temporal regimes. The common shock can be interpolated from neighbouring sites, whereas local temporal dynamics are obtained through the posterior predictive distribution of the spatial partition. The model therefore yields both continuous prediction of the latent environmental process and probabilistic temporal-regime assignment, with uncertainty propagated jointly from covariate effects, spatial interpolation, regime allocation, temporal evolution, and measurement error.

Replicated simulations examine how regime separation and geographic organisation affect clustering recovery, how regime uncertainty propagates into spatial prediction, and what changes when individual model components are removed. We then analyse daily particulate matter with aerodynamic diameter up to $10\,\mu$m (PM10) measurements from 2,989 European monitoring locations in 2024, with prediction at 5,297 unmonitored land locations. The application examines residual temporal heterogeneity, uncertainty in the inferred regimes, and the extent to which those regimes and spatially shared fluctuations provide complementary descriptions of the monitoring network.

The remainder of the paper is organised as follows. Section~\ref{sec:model} introduces the hierarchical model, spatial partition prior, posterior-inference strategy and prediction at unobserved sites. Sections~\ref{sec:simulation} and~\ref{sec:application} report the simulation study and European PM10 analysis, respectively, and  Section~\ref{sec:discussion} concludes. Detailed posterior computation, implementation, and computational benchmarks are provided in the Supplementary Material.

\section{The DCS-sPPM model}
\label{sec:model}

In this section, we introduce our modelling framework for geo-referenced time series, designed to jointly characterise spatial dependence and heterogeneity in local temporal dynamics. Let $y_t(\bs{s}_i)$ be the response at time $t=1,\ldots,T$ and location $\bs{s}_i$, for $i=1,\ldots,n$. Let $\bs{x}_{t}(\bs{s}_i)\in\mathbb{R}^p$ be the vector containing an intercept and covariates that may vary over both location and time. The observation model is
\begin{equation}
\label{eq:observation}
  y_t(\bs{s}_i)
  =
  \bs{x}_{t}(\bs{s}_i)^{\top}\bs{\beta}
  + w_t(\bs{s}_i)
  + \epsilon_t(\bs{s}_i),
  \qquad
  \epsilon_t(\bs{s}_i)\overset{\mathrm{iid}}{\sim}\mathcal{N}(0,\sigma^2),
\end{equation}
where $\bs{\beta}$ is a global regression vector and $w_t(\bs{s}_i)$ is a latent residual process. 

The residual process in \eqref{eq:observation} evolves according to
\begin{equation}
\label{eq:state}
  w_t(\bs{s}_i)
  =
  \phi_i w_{t-1}(\bs{s}_i)
  + u_t(\bs{s}_i)
  + \eta_t(\bs{s}_i),
  \qquad
  \eta_t(\bs{s}_i)\overset{\mathrm{ind}}{\sim}\mathcal{N}(0,\tau_i^2),
\end{equation}
for $t=2,\ldots,T$. Here $\phi_i\in(0,1)$ measures local temporal persistence, $\tau_i^2$ is the variance of the local innovation, and $u_t(\bs{s}_i)$ is a spatially structured common shock at location $\bs{s}_i$. We use the initialisation
\begin{equation}
\label{eq:initial_state}
  w_1(\bs{s}_i)\mid\phi_i,\tau_i^2
  \sim
  \mathcal{N}\!\left(0,\frac{\tau_i^2}{1-\phi_i^2}\right).
\end{equation}
If the common-shock term were absent, the local process would reduce to an AR(1) model with innovation variance $\tau_i^2$, whose stationary variance is $\tau_i^2/(1-\phi_i^2)$. Equation~\eqref{eq:initial_state} uses this stationary distribution to initialise $w_1(\bs{s}_i)$. Because the process is additionally driven by $u_t(\bs{s}_i)$ for $t\geq2$, however, this is not the stationary marginal distribution of the full latent process. The choice of this initialisation preserves the cluster-specific interpretation of the initial state while avoiding the need to introduce a common-shock field at $t=1$. The common shock enters in \eqref{eq:state} only for $t\geq2$; we set $u_1(\bs{s}_i)=0$ for notational convenience. This conditionally Gaussian state-space representation is closely related to standard Bayesian dynamic and spatio-temporal process formulations \citep{gelfand2005spatial,west2006bayesian,durbin2012time}.

For $t\geq2$, the one-step innovation is
\begin{equation*}
  e_t(\bs{s}_i)=w_t(\bs{s}_i)-\phi_iw_{t-1}(\bs{s}_i)
  =u_t(\bs{s}_i)+\eta_t(\bs{s}_i).
\end{equation*}
Under the parent Gaussian-process specification, let $R_{ij}(\bs{\zeta})=R(h_{ij};\bs{\zeta})$ denote the spatial correlation between sites $i$ and $j$, where $\bs{\zeta}$ is the correlation-parameter vector and $h_{ij}=\lVert\bs{s}_i-\bs{s}_j\rVert$. Then
\begin{equation}\label{eq:cov_decomp}
  \Cov\{e_t(\bs{s}_i),e_t(\bs{s}_j)\}
  =\sigma_u^2R_{ij}(\bs{\zeta})+\mathbb{I}(i=j)\tau_i^2.
\end{equation}
Thus, off-diagonal innovation dependence is explained entirely by the common shock and diagonal excess variation by the local regime. The covariance decomposition in \eqref{eq:cov_decomp} is the formal sense in which the two sources of variability are decoupled. The decomposition also determines how the fitted model should be interpreted. The common-shock process describes residual variation that is spatially shared at a given time, whereas $(\phi_i,\tau_i^2)$ describes how site-specific departures evolve after accounting for that shared component. Consequently, the resulting clusters should be interpreted as \emph{local temporal regimes}.

Conditional on the common shocks, the latent trajectories are independent across sites. Conversely, conditional on the latent trajectories and temporal parameters, the common shocks are independent across times.

\subsection{Spatial common shocks and their NNGP representation}
\label{sec:nngp}

Conceptually, let the common shocks $\bs{u}_t=(u_t(\bs{s}_1),\ldots,u_t(\bs{s}_n))^\top$ be independent replicates of a mean-zero Gaussian process,
\begin{equation}\label{eq:u_joint_dist}
  \bs{u}_t\mid\bs{\zeta},\sigma_u^2
  \stackrel{iid}{\sim}
  \mathcal{N}_n(\bs{0},\sigma_u^2\bs{R}(\bs{\zeta})),
  \qquad t=2,\ldots,T,
\end{equation}
where $\bs{R}(\bs{\zeta})$ is the parent correlation matrix with entries $R_{ij}(\bs{\zeta})$, and $\sigma_u^2$ scales the common-shock covariance.

Direct operations on the dense $n\times n$ correlation matrix become computationally demanding for large monitoring networks. We therefore use an NNGP representation \citep{datta2016hierarchical,finley2019efficient} for \eqref{eq:u_joint_dist}, which belongs to the broader class of sparse Vecchia approximations \citep{vecchia1988estimation,katzfuss2021}. For a fixed ordering of the sites, let $N(i)\subset\{1,\ldots,i-1\}$ contain the $\min(m,i-1)$ nearest preceding neighbours of site $i$, with $N(1)=\varnothing$. The approximation is defined by
\begin{equation*}
  u_t(\bs{s}_i)\mid\bs{u}_t\{N(i)\}
  \sim
  \mathcal{N}\!\left(
    \bs{b}_i(\bs{\zeta})^\top\bs{u}_t\{N(i)\},
    \sigma_u^2\delta_i(\bs{\zeta})
  \right),
\end{equation*}
where
\begin{align*}
  \bs{b}_i(\bs{\zeta})
  &=
  \bs{R}_{N(i),N(i)}(\bs{\zeta})^{-1}\bs{R}_{N(i),i}(\bs{\zeta}),\\
  \delta_i(\bs{\zeta})
  &=
  1-\bs{R}_{i,N(i)}(\bs{\zeta})
  \bs{R}_{N(i),N(i)}(\bs{\zeta})^{-1}
  \bs{R}_{N(i),i}(\bs{\zeta}).
\end{align*}
The resulting precision matrix is
\begin{equation*}
\label{eq:nngp_precision}
  \bs{Q}_u(\bs{\zeta},\sigma_u^2)
  =
  \frac{1}{\sigma_u^2}
  \bs{B}(\bs{\zeta})^\top
  \bs{\Delta}(\bs{\zeta})^{-1}
  \bs{B}(\bs{\zeta}),
\end{equation*}
where $\bs{B}(\bs{\zeta})$ is a sparse lower-triangular matrix with unit diagonal and entries $-\bs{b}_i(\bs{\zeta})$ in the columns corresponding to $N(i)$, while $\bs{\Delta}(\bs{\zeta})=\diag\{\delta_1(\bs{\zeta}),\ldots,\delta_n(\bs{\zeta})\}$. Thus, each row of $\bs{B}(\bs{\zeta})$ contains at most $m$ nonzero off-diagonal entries.

The conditional construction also gives a direct innovation-form likelihood. Let
\begin{equation*}
  r_{ti}(\bs{\zeta})=u_t(\bs{s}_i)-
  \bs{b}_i(\bs{\zeta})^\top\bs{u}_t\{N(i)\}.
\end{equation*}
Then, up to a constant,
\begin{align}
 \log p(\bs{u}_{2:T}\mid\bs{\zeta},\sigma_u^2)
 ={}&-\frac{n(T-1)}{2}\log\sigma_u^2
 -\frac{T-1}{2}\sum_{i=1}^n\log\delta_i(\bs{\zeta})-\frac{1}{2\sigma_u^2}\sum_{t=2}^T\sum_{i=1}^n
 \frac{r_{ti}(\bs{\zeta})^2}{\delta_i(\bs{\zeta})}.
 \label{eq:nngp_likelihood}
\end{align}
No dense $n\times n$ covariance matrix is formed. Constructing the neighbour coefficients requires systems of order at most $m$, and evaluating equation~\eqref{eq:nngp_likelihood} is linear in $nT$ for fixed $m$.

\subsection{Clustering of spatially coherent temporal regimes}
\label{sec:clustering}

Locations are clustered through the local temporal parameter
\begin{equation*}
  \bs{\theta}_i=(\phi_i,\tau_i^2).
\end{equation*}
This parameter-driven definition follows the idea of clustering spatially referenced time series according to interpretable features of their underlying temporal dynamics rather than their raw trajectories \citep{aiello2026bayesian}. Let $c_i\in\{1,\ldots,K\}$ denote the allocation of site $i$, let $S_k=\{i:c_i=k\}$, and write $\rho_n=\{S_1,\ldots,S_K\}$. Sites in cluster $k$ share
\begin{equation*}
  (\phi_i,\tau_i^2)=(\phi_k^*,\tau_k^{2*}),
  \qquad i\in S_k.
\end{equation*}

The partition prior is the sPPM of \citet{page2016spatial},
\begin{equation*}
\label{eq:sppm}
  p(\rho_n\mid\bs{s}_1,\ldots,\bs{s}_n)
  \propto
  \prod_{k=1}^K C(S_k)g(\bs{s}_k^*),
\end{equation*}
where $C(S_k)$ is a cohesion and $g(\bs{s}_k^*)$ is a spatial similarity function. We use the Dirichlet process (DP) cohesion $C(S_k)=M\Gamma(|S_k|)$, where $M$ is the mass parameter. The spatial similarity is defined as the marginal likelihood of the locations in a cluster under a bivariate Gaussian model with a conjugate Normal--inverse-Wishart prior, so that spatially compact groups receive larger prior weight. More explicitly, 
\begin{equation*}
  g(\bs{s}_k^*)=
  \int\prod_{i\in S_k}q(\bs{s}_i\mid\bs{m}_k,\bs{V}_k)
  q(\bs{m}_k,\bs{V}_k)\,d\bs{m}_k\,d\bs{V}_k.
  \label{eq:similarity}
\end{equation*}
where
\begin{align*}
  \bs{s}_i\mid\bs{m}_k,\bs{V}_k&\sim
  \mathcal{N}_2(\bs{m}_k,\bs{V}_k),\qquad i\in S_k,\\
  \bs{m}_k\mid\bs{V}_k&\sim
  \mathcal{N}_2(\bs{m}_0,\bs{V}_k/\lambda_0),\\
  \bs{V}_k&\sim\mathrm{IW}(\nu_0,\bs{\Lambda}_0).
\end{align*}

Writing $n_k=|S_k|$, $\bar{\bs{s}}_k$ for the cluster centroid,
$\lambda_k=\lambda_0+n_k$, $\nu_k=\nu_0+n_k$, and
\begin{equation*}
 \bs{\Lambda}_k=\bs{\Lambda}_0+
 \sum_{i\in S_k}(\bs{s}_i-\bar{\bs{s}}_k)(\bs{s}_i-\bar{\bs{s}}_k)^\top+
 \frac{\lambda_0n_k}{\lambda_k}
 (\bar{\bs{s}}_k-\bs{m}_0)(\bar{\bs{s}}_k-\bs{m}_0)^\top,
\end{equation*}
the closed form is
\begin{equation*}
 g(\bs{s}_k^*)=
 \pi^{-n_k}
 \left(\frac{\lambda_0}{\lambda_k}\right)
 \frac{|\bs{\Lambda}_0|^{\nu_0/2}}
 {|\bs{\Lambda}_k|^{\nu_k/2}}
 \frac{\Gamma_2(\nu_k/2)}{\Gamma_2(\nu_0/2)},
 \label{eq:similarity_closed}
\end{equation*}
under the stated inverse-Wishart convention.

\begin{figure}[t]
    \centering
\vspace{1.5em}
\begin{overpic}[width=\textwidth]{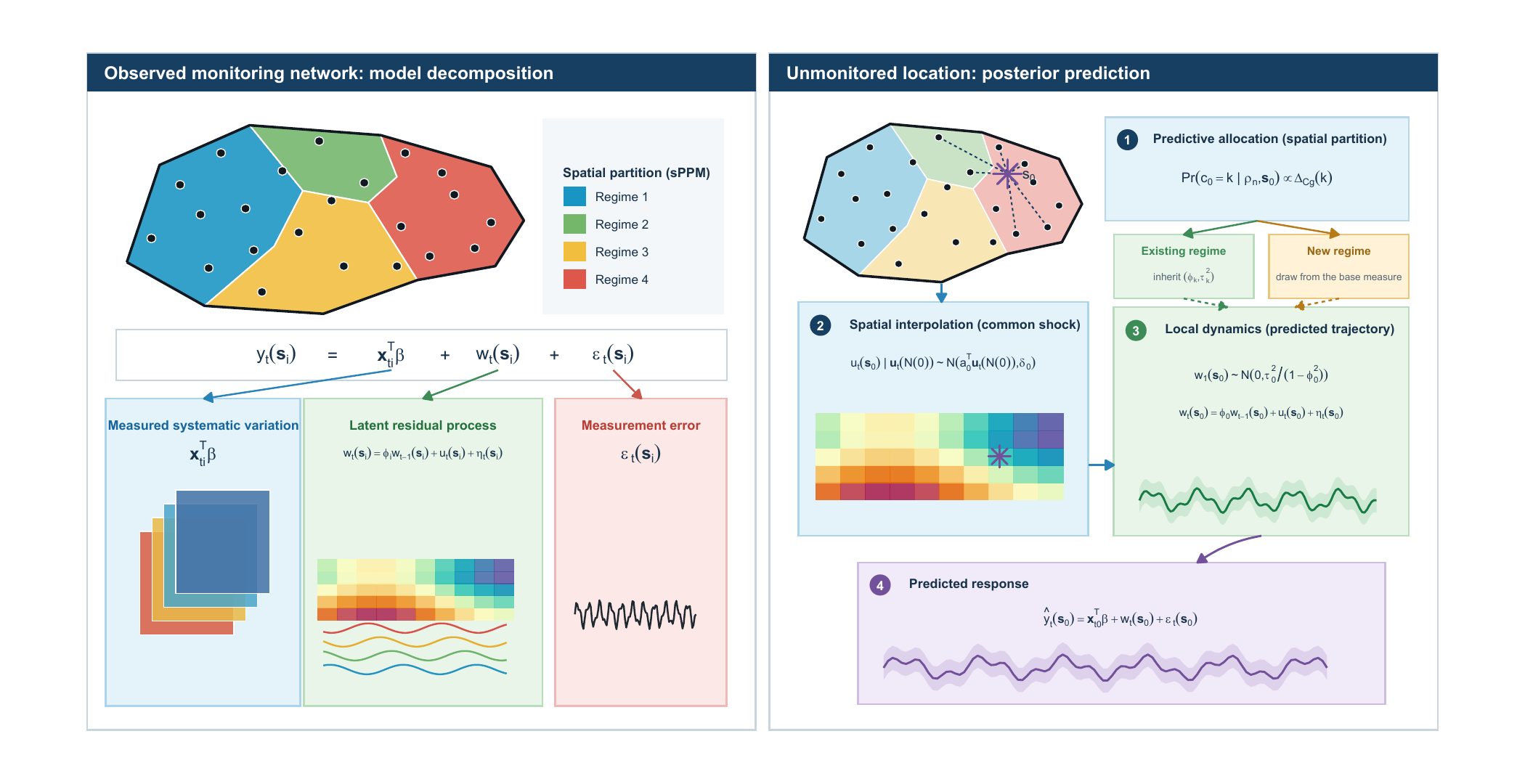}
\put(1,52){\jrsscpanellabel{A}}
\put(51,52){\jrsscpanellabel{B}}
\end{overpic}
    \caption{
    Conceptual representation of the DCS-sPPM decomposition and prediction mechanism.
    }
    \label{fig:dcs_scheme}
\end{figure}

\subsection{Prior specification and posterior inference}
\label{sec:priors_inference}

We assign a multivariate Gaussian prior to the regression coefficients, $\bs{\beta}\sim\mathcal{N}_p(\bs{0},\bs{\Xi})$, with diagonal covariance $\bs{\Xi}=\diag(\xi_1^2,\ldots,\xi_p^2)$ and independent $\xi_l^2\sim\mathrm{IG}(a_\xi,b_\xi)$, $l=1,\ldots,p$. The measurement-error variance follows $\sigma^2\sim\mathrm{IG}(a_\sigma,b_\sigma)$. For the cluster-specific temporal parameters, we assume independent $\phi_k^*\sim\mathrm{Beta}(a_\phi,b_\phi)$ and $\tau_k^{2*}\sim\mathrm{IG}(a_\tau,b_\tau)$. For the implementation used in the simulations and European application, we employ the exponential correlation $R(h;\kappa)=\exp(-h/\kappa)$, equivalent to a Mat\'ern correlation with smoothness $1/2$. The correlation-parameter vector therefore reduces to the scalar spatial range $\kappa$. The common-shock variance is assigned $\sigma_u^2\sim\mathrm{IG}(a_{\sigma_u},b_{\sigma_u})$, while the spatial range follows $\kappa\sim\mathrm{Ga}(a_\kappa,b_\kappa)$. We parameterise $\mathrm{IG}(a,b)$ through the density $p(v)\propto v^{-a-1}\exp(-b/v)$ for $v>0$, and use a shape--rate parameterisation for the Gamma distribution. Numerical hyperparameter values are reported in Section~\ref{sec:application_specification} for the European analysis and Supplementary Section~\ref{supp:additional_simulation} for the simulation experiments.

Posterior inference is performed with a hybrid Gibbs and Metropolis--Hastings sampler that exploits the conditional structure of the model. Local latent trajectories are sampled through Gaussian updates based on tridiagonal precision matrices, while cluster allocations are updated sitewise conditional on the current latent trajectories and cluster-specific temporal parameters. The spatial covariance parameters are updated through a collapsed Metropolis--Hastings step, after which the common-shock process is sampled using the sparse NNGP precision structure. For the European application, a representative partition is obtained by minimising posterior expected variation-of-information (VI) loss \citep{meilua2007,wade2018,dahl2022}. Full conditional distributions, implementation details and computational complexity are provided in Supplementary Section~\ref{supp:posterior}.

\subsection{Spatio-temporal prediction}
\label{sec:prediction}

Consider an unobserved location $\bs{s}_0$. For each retained posterior draw, prediction proceeds by sampling a temporal regime, interpolating the common shock, and recursively generating the local state. This leads to two distinct predictive targets. The first is \emph{regime prediction}, which quantifies the local temporal dynamics that are plausible at $\bs{s}_0$. The second is \emph{process prediction}, which reconstructs the latent environmental trajectory and, if desired, the corresponding observed response.

Let $S_1,\ldots,S_K$ be the current occupied clusters. Conditional on the current posterior partition, the predictive allocation of $\bs{s}_0$ is determined by the following cohesion and spatial-similarity ratios. Because no response is observed at the new location, no additional response-likelihood contribution enters these weights::
\begin{equation}
\label{eq:predictive_allocation}
  \Pr(c_0=k\mid\rho_n,\bs{s}_0)\propto
  \begin{cases}
  \dfrac{C(S_k\cup\{0\})g(\bs{s}_k^*\cup\{\bs{s}_0\})}
        {C(S_k)g(\bs{s}_k^*)}, & k=1,\ldots,K,\\[8pt]
  C(\{0\})g(\{\bs{s}_0\}), & k=K+1.
  \end{cases}
\end{equation}
An existing assignment inherits $(\phi_k^*,\tau_k^{2*})$; a new cluster draws these parameters from the base measure. This predictive allocation follows the spatial product-partition construction of \citet{page2016spatial}. More generally, similarity-weighted predictive allocation of a new unit is closely related to covariate-dependent product partition models, in which available covariate information modifies the probability of joining existing clusters \citep{muller2011ppm}.

In addition to supporting prediction of the latent trajectory and response, the predictive partition also yields label-invariant summaries of local temporal-regime characteristics at unobserved locations, such as posterior predictive means of $\phi$ and $\tau^2$ and measures of allocation uncertainty.

For prediction, we retain the NNGP ordering of the observed locations and place the unmonitored location $\bs{s}_0$ after them. Thus, all observed locations precede $\bs{s}_0$ and are eligible to enter its conditioning set. We take $N(0)$ to contain its $m$ nearest observed neighbours. The resulting NNGP conditional distribution is
\begin{equation}
\label{eq:u_prediction}
  \widetilde{u}_t(\bs{s}_0)\mid\bs{u}_t,\kappa,\sigma_u^2
  \sim
  \mathcal{N}\!\left(
    \bs{b}_0(\kappa)^\top\bs{u}_t\{N(0)\},
    \sigma_u^2\delta_0(\kappa)
  \right),
\end{equation}
for $t=2,\ldots,T$. Equation~\eqref{eq:u_prediction} provides the continuous spatial component of prediction at the new location. Conditional on the sampled regime, the local latent trajectory is then generated recursively as
\begin{equation}
\label{eq:w_prediction}
  \widetilde{w}_t(\bs{s}_0)
  =
  \phi_0\widetilde{w}_{t-1}(\bs{s}_0)
  +\widetilde{u}_t(\bs{s}_0)
  +\widetilde{\eta}_t(\bs{s}_0),
  \qquad
  \widetilde{\eta}_t(\bs{s}_0)\sim\mathcal{N}(0,\tau_0^2),
  \quad t=2,\ldots,T.
\end{equation}
The initial state satisfies $\widetilde{w}_1(\bs{s}_0) \sim \mathcal{N}\!\left(0,\frac{\tau_0^2}{1-\phi_0^2}\right)$. Finally, equation~\eqref{eq:w_prediction} is combined with the global regression mean to obtain a predictive response,
\begin{equation}
\label{eq:y_prediction}
  \widetilde{y}_t(\bs{s}_0)
  =
  \bs{x}_{t0}^\top\bs{\beta}
  +\widetilde{w}_t(\bs{s}_0)
  +\widetilde{\epsilon}_t(\bs{s}_0),
  \qquad
  \widetilde{\epsilon}_t(\bs{s}_0)\sim\mathcal{N}(0,\sigma^2).
\end{equation}
Here $\bs{x}_{t0}=\bs{x}_t(\bs{s}_0)$ denotes the supplied covariate vector at the prediction location. Equation~\eqref{eq:y_prediction} defines the posterior predictive distribution of the response by combining systematic covariate variation, the predicted latent trajectory and measurement error.

The resulting hierarchy separates two distinct sources of spatial information: the common-shock process captures continuously varying dependence across locations, whereas the sPPM induces a discrete spatial organisation of local temporal dynamics through $(\phi_i,\tau_i^2)$ after adjustment for the global covariate effect and the common shock. Figure~\ref{fig:dcs_scheme} illustrates this decomposition and how the two components are extended to an unmonitored location $\bs{s}_0$.

\section{Simulation study}
\label{sec:simulation}

The simulation study examines temporal-regime recovery under varying separation and geographic organisation (Section~\ref{sec:simulation_recovery}), prediction and regime uncertainty at unmonitored locations (Section~\ref{sec:simulation_prediction}), and the contribution of the individual model components through a four-model comparison (Supplementary Section~\ref{supp:comparative_simulation}).

\subsection{Temporal-regime recovery under varying spatial settings}
\label{sec:simulation_recovery}

The main clustering experiment used a $12\times12$ regular lattice on $[0,5]^2$, hence $n=144$, and $T=365$ time points. Three balanced regimes were arranged as vertical strips (\emph{regular}), as contiguous regions separated by nonlinear boundaries (\emph{irregular}), or by a spatially random balanced allocation (\emph{random}). For the irregular case labels were assigned following the ordering score $\Pi(s_{i1},s_{i2})=s_{i1}-0.85\sin(2\pi s_{i2}/5)-0.20\sin(4\pi s_{i2}/5)$,
and its two empirical tertiles defined the regions. Regime labels were permuted across regions independently in every replicate so that results could not depend on a fixed label order.

We also considered three levels of separation among the temporal regimes. Writing $\bs{\theta}^{*}_k=(\phi^{*}_k,\tau^{2*}_k)^\top$ for the parameters of cluster $k$, the \emph{similar} configuration used $\bs{\theta}^{*}_1=(0.45,0.30)^\top$, $\bs{\theta}^{*}_2=(0.55,0.35)^\top$, and $\bs{\theta}^{*}_3=(0.65,0.40)^\top$. The \emph{intermediate} configuration increased the separation in both persistence and innovation variability, with $\bs{\theta}^{*}_1=(0.25,0.20)^\top$, $\bs{\theta}^{*}_2=(0.55,0.35)^\top$, and $\bs{\theta}^{*}_3=(0.80,0.55)^\top$, while the \emph{different} configuration used the more clearly separated values $\bs{\theta}^{*}_1=(0.10,0.10)^\top$, $\bs{\theta}^{*}_2=(0.55,0.35)^\top$, and $\bs{\theta}^{*}_3=(0.90,0.80)^\top$. Across all configurations, the remaining parameters were fixed at $\bs{\beta}=(1,1.25,-0.8)^\top$, $\sigma^2=0.08$, $\sigma_u^2=0.35$, and $\kappa=1.5$. Two space--time covariates were constructed as $x_{it,1}=\sin(2\pi t/24)+0.35s_{i1}+e_{it,1}$ and $x_{it,2}=\cos(2\pi t/17)s_{i2}+e_{it,2}$, combining periodic temporal variation, standardised spatial coordinates, and Gaussian noise. The errors were generated independently across sites, times, and covariates, with $e_{it,1}\sim\mathcal{N}(0,0.55^2)$ and $e_{it,2}\sim\mathcal{N}(0,0.65^2)$, and the resulting covariates were standardised before use. Data generation and fitting used the same NNGP specification with $m=10$ neighbours. The experiment therefore evaluates inferential recovery under correct specification rather than NNGP approximation error. Each of the nine scenarios was replicated 100 times, using chains of 8,000 iterations, with 4,000 discarded as burn-in and no thinning.

For the simulation experiments, a representative partition was selected from the retained draws by minimising the squared discrepancy between its pairwise co-membership matrix and the posterior co-clustering matrix. Recovery was evaluated using the adjusted Rand index (ARI) \citep{hubert1985comparing}, VI, pairwise sensitivity and specificity, Brier loss \citep{brier1950verification} for posterior co-clustering probabilities, exact recovery up to label permutation, and the posterior number of clusters. Cluster-specific parameters were matched to the data-generating regimes before summarisation.

Table~\ref{tab:clustering_recovery} summarises the central empirical result: the latent temporal partition is recovered almost perfectly once the regimes are moderately or clearly separated. Recovery is more difficult when regimes overlap substantially, especially when their labels are spatially random, because neither the temporal likelihood nor the spatial prior then provides strong discriminatory information. Conversely, spatially coherent regimes remain well recovered even under weak separation. Exact recovery is deliberately stringent and may therefore remain low even when the representative partition closely matches the data-generating structure according to the ARI and VI. The co-clustering Brier loss confirms the same pattern by comparing the posterior probability that each pair of sites belongs to the same cluster with their true co-membership status.

\begin{table}[t]
\centering
\footnotesize
\caption{Clustering recovery over 100 replications. ``Exact'' is the proportion of representative partitions equal to the data-generating partition up to label permutation.}
\label{tab:clustering_recovery}
\begin{tabular}{llrrrrrrrr}
\toprule
Geography & Separation & ARI & VI & Sens. & Spec. & Brier & Exact & $E(K\mid y)$ & Modal $K=3$ \\
\midrule
Regular & Similar      & 0.9672 & 0.0983 & 0.9780 & 0.9892 & 0.0114 & 0.25 & 3.097 & 0.98 \\
Regular & Intermediate & 0.9995 & 0.0016 & 0.9993 & 1.0000 & 0.0002 & 0.96 & 3.111 & 0.96 \\
Regular & Different    & 0.9998 & 0.0007 & 0.9997 & 1.0000 & 0.0001 & 0.98 & 3.072 & 0.97 \\
Irregular & Similar      & 0.9239 & 0.2077 & 0.9498 & 0.9745 & 0.0246 & 0.01 & 3.127 & 1.00 \\
Irregular & Intermediate & 0.9999 & 0.0003 & 0.9999 & 1.0000 & 0.0002 & 0.99 & 3.150 & 0.99 \\
Irregular & Different    & 1.0000 & 0.0000 & 1.0000 & 1.0000 & 0.0003 & 1.00 & 3.121 & 0.98 \\
Random & Similar      & 0.6537 & 0.7988 & 0.7677 & 0.8862 & 0.0993 & 0.00 & 3.443 & 0.81 \\
Random & Intermediate & 0.9913 & 0.0165 & 0.9889 & 1.0000 & 0.0029 & 0.88 & 3.454 & 0.82 \\
Random & Different    & 0.9954 & 0.0087 & 0.9941 & 1.0000 & 0.0021 & 0.94 & 3.368 & 0.89 \\
\bottomrule
\end{tabular}
\end{table}

The posterior mean number of clusters remains close to three in the spatially coherent scenarios. It is slightly larger under random labels because the sPPM compromises between a non-spatial true allocation and a prior favouring compactness. The fact that the random intermediate and different cases still reach high ARI values demonstrates that the likelihood can overcome this preference when temporal regimes are distinguishable. The random-similar scenario instead illustrates the difficulty of recovering weakly separated regimes when geography is uninformative. This distinction is important for environmental interpretation: a spatial partition prior can improve inference for weakly separated regimes only when geographic organisation is itself a scientifically plausible feature of those regimes.

Table~\ref{tab:parameter_recovery} assesses recovery of the parameters used to generate the simulated data. The true values are reported alongside the recovery measures: global parameters are fixed across all scenarios, whereas some cluster-specific parameters vary with regime separation but not with geographic configuration. For each parameter and each of the nine combinations of geographic configuration and regime separation, we first summarise relative bias, root mean squared error (RMSE), empirical 95\% coverage, and within-chain effective sample size (ESS) over the 100 replications. The table then reports the minimum and maximum of these summaries. Before evaluating the cluster-specific parameters, estimated regimes were matched to the data-generating regimes to remove arbitrary label permutations. The global regression and spatial covariance parameters are recovered consistently across scenarios. Recovery of the regime-specific persistence parameters is less uniform when the regimes overlap, reflecting uncertainty in both the partition and the associated parameters, whereas the innovation variances are more stable.

\begin{table}[t]
\centering
\footnotesize
\caption{Recovery of the continuous data-generating parameters. Performance entries report ranges across the nine combinations of geographic configuration and regime separation, with 100 replications per scenario. Where three true values are reported, they correspond to the similar, intermediate, and different separation configurations, respectively; these values are unchanged across geographic configurations. RB denotes relative bias, RMSE root mean squared error, and ESS effective sample size. Cluster-specific parameters were matched to the true regimes before evaluation.}
\label{tab:parameter_recovery}
\begin{tabular}{lcrrrr}
\toprule
Parameter & True value(s) & RB & RMSE & 95\% coverage & Median ESS\\
\midrule
$\beta_0$ & $1$ & $[-0.0030,0.0013]$ & $[0.0093,0.0263]$ & $[0.93,0.96]$ & $[73,260]$\\
$\beta_1$ & $1.25$ & $[-0.0004,0.0011]$ & $[0.0044,0.0059]$ & $[0.90,0.97]$ & $[693,1395]$\\
$\beta_2$ & $-0.8$ & $[-0.0006,0.0005]$ & $[0.0040,0.0045]$ & $[0.93,0.97]$ & $[894,1634]$\\
$\sigma^2$ & $0.08$ & $[-0.0251,0.0044]$ & $[0.0071,0.0087]$ & $[0.89,0.96]$ & $[65,90]$\\
$\kappa$ & $1.5$ & $[-0.0012,0.0103]$ & $[0.0602,0.0642]$ & $[0.92,0.96]$ & $[177,231]$\\
$\sigma_u^2$ & $0.35$ & $[-0.0023,0.0026]$ & $[0.0081,0.0090]$ & $[0.91,0.97]$ & $[267,314]$\\
$\sigma_u^2/\kappa$ & $0.2333$ & $[-0.0076,0.0018]$ & $[0.0094,0.0106]$ & $[0.88,0.95]$ & $[113,170]$\\
$\phi_1$ & $0.45/0.25/0.10$ & $[-0.0158,0.0305]$ & $[0.0089,0.0176]$ & $[0.61,0.98]$ & $[90,166]$\\
$\phi_2$ & $0.55/0.55/0.55$ & $[-0.0033,0.0009]$ & $[0.0076,0.0105]$ & $[0.89,0.98]$ & $[130,163]$\\
$\phi_3$ & $0.65/0.80/0.90$ & $[-0.0181,-0.0002]$ & $[0.0033,0.0149]$ & $[0.65,0.94]$ & $[144,468]$\\
$\tau_1^2$ & $0.30/0.20/0.10$ & $[-0.0021,0.0276]$ & $[0.0086,0.0142]$ & $[0.87,0.96]$ & $[65,108]$\\
$\tau_2^2$ & $0.35/0.35/0.35$ & $[-0.0014,0.0084]$ & $[0.0106,0.0134]$ & $[0.88,0.96]$ & $[82,113]$\\
$\tau_3^2$ & $0.40/0.55/0.80$ & $[-0.0103,0.0052]$ & $[0.0134,0.0185]$ & $[0.92,0.97]$ & $[106,125]$\\
\bottomrule
\end{tabular}
\end{table}

Mean elapsed time per 8,000-iteration fit was approximately 300 seconds across the nine cells on an Intel(R) Core(TM) i7-10750H CPU @ 2.600GHz machine.

\subsection{Prediction and regime uncertainty at unmonitored locations}
\label{sec:simulation_prediction}

The prediction experiment retained the same $n=144$ training design and nine scenarios but used $T=100$ to allow evaluation at every time point in 900 independently fitted data sets. The evaluation set comprised 100 interior grid locations and 40 exterior locations, with ten placed along each edge at a perpendicular distance of 0.35 spatial units beyond the training square $[0,5]^2$ (7\% of its side length), to assess interpolation and controlled short-range extrapolation. Training and prediction processes were generated jointly, after which all responses at prediction sites were removed. Predictive performance was evaluated for the regime assigned to each held-out location and, for each process, at every combination of prediction location and time point. Each fit used 8,000 iterations with 4,000 discarded. Parameter draws were retained without thinning, whereas predictive draws were generated every tenth retained iteration, giving 400 draws per fit; predictive intervals and CRPS were calculated from these draws.

Table~\ref{tab:prediction_processes} evaluates the three stages of the predictive construction in equations~\eqref{eq:u_prediction}--\eqref{eq:y_prediction}: the interpolated common shock $u_{t}(\bs{s})$, the recursively generated latent state $w_{t}(\bs{s})$, and the resulting response $y_{t}(\bs{s})$. Empirical average prediction errors, reported in Supplementary Section~\ref{supp:additional_simulation}, were close to zero for all three processes, indicating no systematic tendency to over- or underpredict under the correctly specified simulation design. Common-shock RMSE is nearly invariant across regime scenarios because $u_{t}(\bs{s})$ is governed by the shared spatial process rather than the partition. The larger RMSE and interval width for $w_{t}(\bs{s})$ and $y_{t}(\bs{s})$ under different regimes are expected: that design includes $\tau_3^2=0.80$, so the target contains substantially more unpredictable local innovation. Coverage nevertheless stays close to 0.95.

\begin{table}[t]
\centering
\footnotesize
\caption{Prediction over 100 replications per scenario and all 140 new sites. Each entry is RMSE / empirical 95\% coverage.}
\label{tab:prediction_processes}
\begin{tabular}{llrrr}
\toprule
Geography & Separation & Common shock $u$ & Latent state $w$ & Response $y$\\
\midrule
Regular & Similar      & 0.364 / 0.947 & 0.847 / 0.946 & 0.892 / 0.946\\
Regular & Intermediate & 0.364 / 0.947 & 0.994 / 0.948 & 1.032 / 0.947\\
Regular & Different    & 0.363 / 0.947 & 1.408 / 0.951 & 1.435 / 0.950\\
Irregular & Similar      & 0.364 / 0.947 & 0.847 / 0.945 & 0.893 / 0.945\\
Irregular & Intermediate & 0.363 / 0.947 & 0.997 / 0.945 & 1.036 / 0.945\\
Irregular & Different    & 0.362 / 0.947 & 1.413 / 0.945 & 1.441 / 0.945\\
Random & Similar      & 0.365 / 0.946 & 0.850 / 0.945 & 0.895 / 0.945\\
Random & Intermediate & 0.362 / 0.946 & 1.013 / 0.944 & 1.052 / 0.944\\
Random & Different    & 0.356 / 0.945 & 1.448 / 0.943 & 1.475 / 0.943\\
\bottomrule
\end{tabular}
\end{table}

Table~\ref{tab:prediction_location_type} separates common-shock prediction at the 100 interior sites from prediction at the 40 exterior sites. Prediction is less precise outside the training domain, as shown by the larger RMSE and wider posterior intervals, whereas empirical coverage remains essentially unchanged. The conditional NNGP distribution therefore reflects the weaker spatial information available for short-range extrapolation by increasing predictive uncertainty.

\begin{table}[t]
\centering
\footnotesize
\caption{Common-shock prediction by location type. Entries report the range across the nine simulation scenarios.}
\label{tab:prediction_location_type}
\begin{tabular}{lrrr}
\toprule
Prediction sites & RMSE range & 95\% coverage range & Interval-width range\\
\midrule
Interpolation & $[0.329,0.338]$ & $[0.946,0.947]$ & $[1.263,1.308]$\\
Extrapolation & $[0.417,0.424]$ & $[0.946,0.947]$ & $[1.602,1.639]$\\
\bottomrule
\end{tabular}
\end{table}

The predictive regime assignment behaves as expected because equation~\eqref{eq:predictive_allocation} can use only coordinates at a new site. Table~\ref{tab:prediction_regimes} shows near-perfect classification for regular-intermediate and different regimes and very good accuracy for irregular regions. In the random scenarios, accuracy is close to the one-third value expected from random classification among three equally represented regimes. The similar-regular design is harder because uncertainty about the temporal parameters propagates into the fitted partition, although its classification accuracy remains at an excellent level. The random-allocation result does not indicate an MCMC failure: under this data-generating mechanism, the regime label of a new site is independent of its location, and therefore cannot be predicted from coordinates alone without observing its response.

Mean runtime was around 130 seconds per 8,000-iteration fit including online prediction. Together, Tables~\ref{tab:prediction_processes} and~\ref{tab:prediction_regimes} separate two questions that should not be conflated: continuous spatial prediction remains calibrated even when regime labels are spatially unpredictable, while classification of a new regime necessarily relies on spatial organisation.

\begin{table}[t]
\centering
\footnotesize
\caption{Predictive regime classification. Entries are hard classification accuracy / posterior probability assigned to the true regime.}
\label{tab:prediction_regimes}
\begin{tabular}{lrrr}
\toprule
Geography & Similar & Intermediate & Different\\
\midrule
Regular   & 0.883 / 0.790 & 1.000 / 0.903 & 1.000 / 0.905\\
Irregular & 0.813 / 0.710 & 0.849 / 0.797 & 0.850 / 0.799\\
Random    & 0.337 / 0.333 & 0.338 / 0.334 & 0.341 / 0.335\\
\bottomrule
\end{tabular}
\end{table}

\subsection{Comparison with reduced model specifications}
\label{sec:simulation_comparison}

We compared four specifications fitted to the same replicated data sets in the nine geography--separation scenarios: the full DCS-sPPM; a non-spatial PPM with the common shock retained; a single-regime model with the common shock retained; and an sPPM without the common shock. These restrictions remove geographic information from the partition prior, heterogeneity in local temporal dynamics, and continuously varying spatial residual dependence, respectively. Each comparison therefore isolates a defining ingredient of the proposed decomposition. Because the data were generated under the full model, this is a controlled structural-ablation study rather than evidence of universal superiority across data-generating mechanisms.

\begin{figure}[t]
\centering
\begin{overpic}[width=\textwidth]{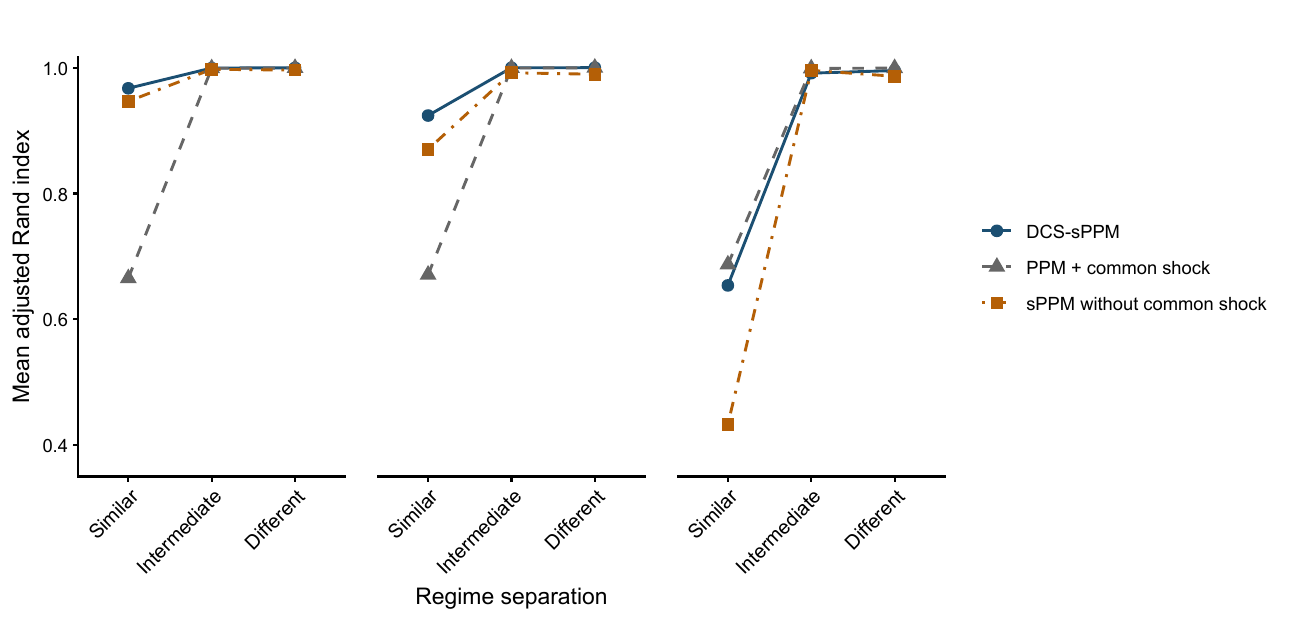}
\put(5.5,47){\jrsscpanellabel{A}}
\put(29,47){\jrsscpanellabel{B}}
\put(52.5,47){\jrsscpanellabel{C}}
\end{overpic}
\caption{Mean adjusted Rand index over 100 replications for (A) regular strips, (B) irregular regions and (C) random allocations, under three levels of regime separation. The plotted specifications are DCS-sPPM, PPM with a common shock, and sPPM without a common shock. The single-regime comparator is omitted because its adjusted Rand index is zero in these three-regime designs.}
\label{fig:comparative_partition_recovery}
\end{figure}

Figure~\ref{fig:comparative_partition_recovery} shows that the spatial partition prior improves recovery of weakly separated regimes when their membership is geographically organised. Under random allocations, the DCS-sPPM offers no corresponding advantage over the non-spatial PPM, while recovery deteriorates further when the common shock is omitted in the weak-separation setting. With greater regime separation, the temporal likelihood becomes sufficiently informative for all three partition models to recover the regimes accurately. These results support using geography as complementary information rather than as a substitute for distinguishable temporal behaviour.

Table~\ref{tab:comparative_prediction} compares prediction under all four models at the same held-out sites and times. DCS-sPPM and PPM with a common shock yield similar continuous predictive performance, showing that spatially informed regime mapping can be obtained without sacrificing the predictive accuracy supplied by the shared spatial process. Omitting the common shock increases latent-state and response prediction errors, whereas imposing a single temporal regime reduces predictive-interval coverage when local dynamics differ across sites. Together with the partition-recovery results, these comparisons support the complementary roles of continuous spatial dependence, heterogeneous temporal dynamics and geographic information in regime allocation. The reported ranges describe variation across scenarios rather than paired effect estimates. Detailed comparative results are provided in Supplementary Section~\ref{supp:comparative_simulation}.

\begin{table}[t]
\centering
\footnotesize
\caption{Ranges of predictive RMSE and empirical 95\% coverage across the nine simulation scenarios, each evaluated over 100 replications at all 140 held-out sites and every time point. Ranges describe scenario-specific summaries, not paired differences between models. A dash denotes a process absent by construction.}
\label{tab:comparative_prediction}
\begin{tabular}{lrrrrrr}
\toprule
& \multicolumn{2}{c}{Common shock $u$} & \multicolumn{2}{c}{Latent state $w$} & \multicolumn{2}{c}{Response $y$}\\
Model & RMSE & Coverage & RMSE & Coverage & RMSE & Coverage\\
\midrule
DCS-sPPM & [0.356,0.365] & [0.945,0.947] & [0.847,1.448] & [0.943,0.951] & [0.892,1.475] & [0.943,0.950]\\
PPM + shock & [0.356,0.365] & [0.946,0.947] & [0.848,1.455] & [0.944,0.946] & [0.893,1.482] & [0.944,0.946]\\
$K=1$ + shock & [0.365,0.422] & [0.926,0.946] & [0.847,1.485] & [0.917,0.943] & [0.892,1.510] & [0.922,0.944]\\
sPPM, no shock & --- & --- & [1.014,1.576] & [0.940,0.948] & [1.051,1.599] & [0.941,0.948]\\
\bottomrule
\end{tabular}
\end{table}

Finally, a standalone benchmark compared runtime and memory requirements for exact Gaussian-process and NNGP calculations as network size increased (Supplementary Section~\ref{sec:nngp_vs_exact}). This benchmark evaluates computational efficiency rather than inferential approximation error and is distinct from the whole-sampler timings.

\section{European PM10 application}
\label{sec:application}

Particulate-matter concentrations arise from processes operating over markedly different spatial and temporal scales. Large-scale meteorological conditions and pollution dynamics can generate fluctuations shared across broad regions, while local emissions, topography, station environment, and other unmeasured factors can produce substantial heterogeneity in persistence and short-term variability among monitoring sites \citep{viana2008source}. A continental monitoring network therefore provides a natural setting in which to distinguish continuously shared spatial variation from local temporal regimes.

The analysis characterises residual temporal behaviour after adjustment for measured meteorological variables and shared spatial fluctuations, identifies where regime membership is well supported or uncertain, and extends both the temporal-regime description and concentration field between monitoring locations. Its applied value is to distinguish similarities in local dynamics from similarities in pollution levels, while showing where the monitoring network supports those distinctions. The regimes are descriptive summaries of residual behaviour, not inferred emissions sources or administrative pollution regions.

\subsection{Data and preprocessing}

The analysis uses daily PM10 observations from European monitoring stations during 2024 \citep{eea_air_quality_2025}. Hourly and subdaily observations were aggregated by monitoring series and day after validity filtering. Series were eligible when observations were available for at least 90\% of the 366 days. Remaining missing days were filled by linear interpolation between available observations, with nearest-endpoint extension where necessary. After temporal and geographic filtering, an audited selection procedure retained one monitoring series per unique spatial location, avoiding duplicate representation of colocated sampling points. The final analysis comprises 2,989 monitoring locations. Further details of data construction and additional exploratory summaries are provided in Supplementary Section~\ref{supp:pm10}.

PM10 concentrations vary substantially over the year, both in their daily average across monitoring stations and in their spread across stations. The corresponding empirical summaries are presented in Supplementary Section~\ref{supp:pm10_data}.

Figure~\ref{fig:europe_exploratory} provides an empirical overview of the resulting monitoring network. The stations are distributed highly unevenly across the study domain, and both annual mean PM10 concentration and station-level lag-one autocorrelation exhibit marked spatial variation. The latter is particularly relevant for our modelling strategy: heterogeneity across monitoring sites concerns not only pollution levels, but also the temporal persistence of the observed series. These descriptive summaries therefore motivate a model that can accommodate both continuously varying spatial structure and location-specific temporal dynamics.

\begin{figure}[t]
\centering
\begin{overpic}[width=\textwidth]{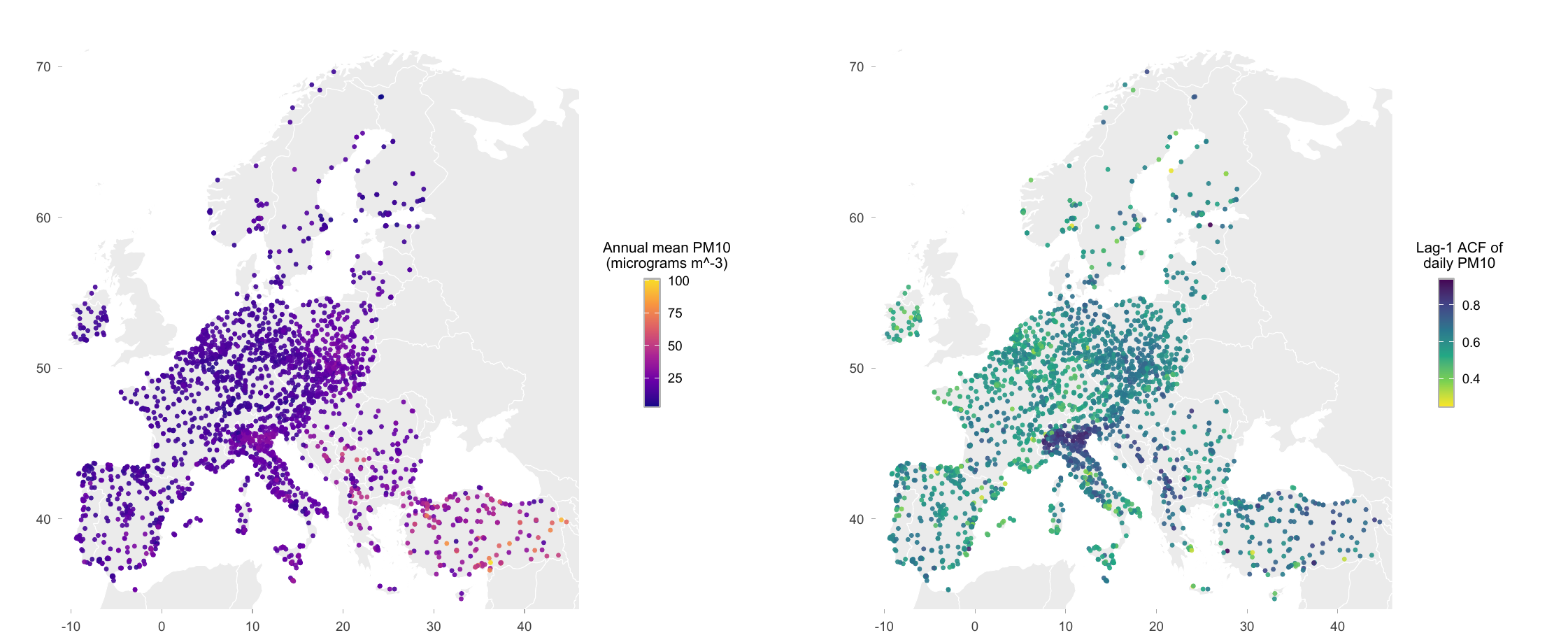}
\put(0.75,39.3){\jrsscpanellabel{A}}
\put(51.3,39.3){\jrsscpanellabel{B}}
\end{overpic}
\caption{Empirical summaries at the European monitoring locations used in the analysis: (A) annual mean PM10 concentration in $\mu\mathrm{g}/\mathrm{m}^3$; (B) lag-one autocorrelation of the daily series.}
\label{fig:europe_exploratory}
\end{figure}

At each prediction location, the daily temperature, precipitation, and wind-speed values were taken from the ERA5 grid cell \citep{c3s_era5_single_levels} associated with the nearest monitoring station.

Candidate prediction locations were formed from a $100\times100$ longitude--latitude grid over the retained domain and intersected with a European land polygon. Open-water cells were excluded, leaving 5,297 prediction locations. For spatial distance calculations, longitude and latitude were converted to approximate planar coordinates, expressed in units of 1,000 km, using an equirectangular transformation with a reference latitude equal to the mean latitude of the monitoring locations.

\subsection{Model specification and computation}
\label{sec:application_specification}

The NNGP used $m=10$ neighbours and max--min ordering \citep{katzfuss2021}. The inverse-Gamma hyperparameters were $(a_\sigma,b_\sigma)=(3,0.40)$, $(a_\xi,b_\xi)=(2,2)$, $(a_\tau,b_\tau)=(2,0.35)$, and $(a_{\sigma_u},b_{\sigma_u})=(3,0.60)$. We used $(a_\phi,b_\phi)=(1,1)$, the shape--rate Gamma prior $(a_\kappa,b_\kappa)=(3,2)$, and DP cohesion mass $M=1$. Coordinates entering the sPPM similarity were standardised. These transformed coordinates affect only cluster compactness; NNGP distances use the separate metric coordinate representation. For the spatial similarity, we used $\lambda_0=1$, $\nu_0=4$, $\bs{\Lambda}_0=\bs{I}_2$, and $\bs{m}_0$ equal to the empirical mean of the standardised monitoring coordinates.

The sampler ran for 30,000 iterations, with 20,000 discarded as burn-in and 10,000 retained, using eight OpenMP threads. Posterior prediction was performed over all 5,297 unmonitored locations and 366 days for the common shock $u_t(\bs{s})$, latent state $w_t(\bs{s})$, and response $y_t(\bs{s})$. The fit required approximately 14 hours, including online prediction; comparative runtimes and computational diagnostics are reported in the Supplementary Material. Predictive temporal-regime summaries over the land grid were obtained by post-processing retained posterior partition output, without rerunning the MCMC sampler.

\subsection{Results}

The global mean component includes temperature, precipitation, and wind speed as meteorological covariates. These variables account for systematic variation in PM10 concentrations before the latent spatial and temporal components are introduced. Consequently, the temporal regimes discussed below describe heterogeneity in persistence and local innovation variability after accounting for both the measured meteorological covariates and the spatially shared common shock.

\subsubsection{Environmental temporal regimes and their uncertainty}

\begin{figure}[t]
\centering
\begin{overpic}[width=\textwidth]{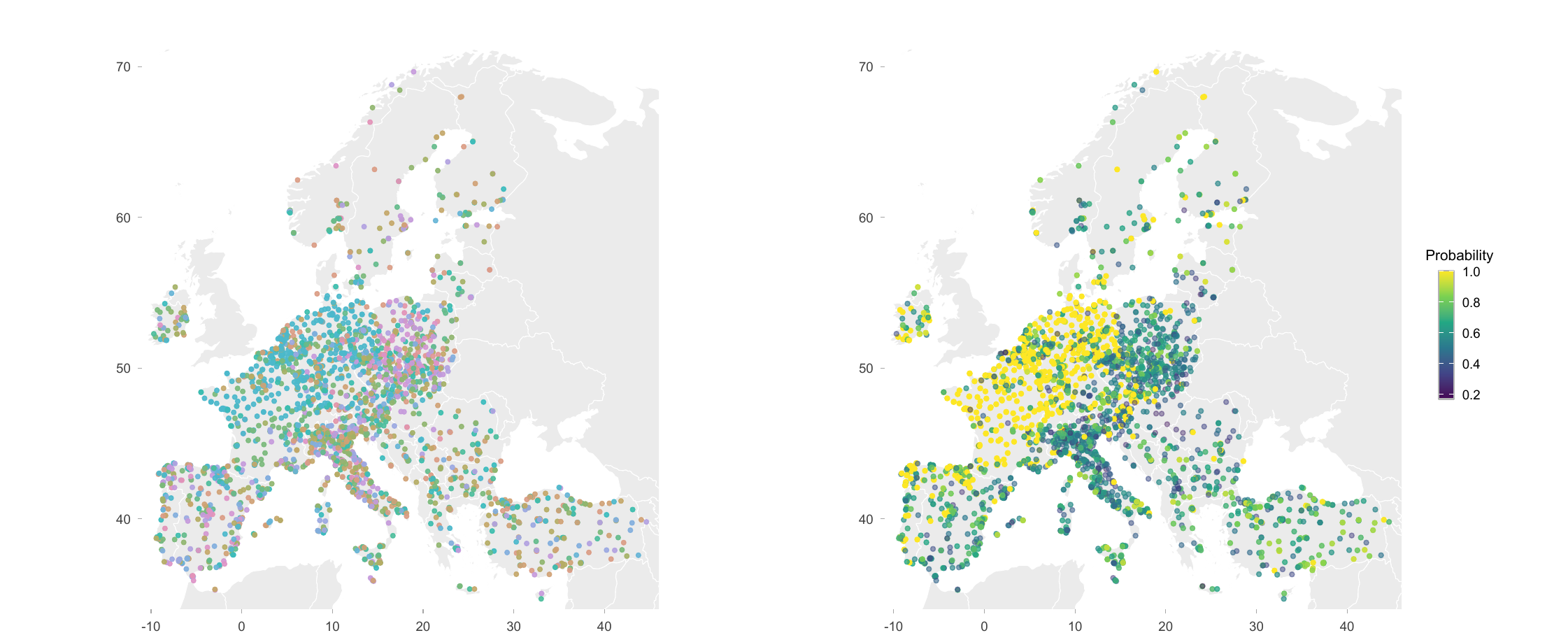}
\put(0.75,39.3){\jrsscpanellabel{A}}
\put(51.3,39.3){\jrsscpanellabel{B}}
\end{overpic}
\caption{Posterior clustering summaries: (A) the VI representative partition, with graph colours reused across groups; (B) mean within-group co-clustering probability. Country borders are geographic reference only and do not enter the partition prior.}
\label{fig:vi_partition}
\end{figure}

European stations show substantial heterogeneity in residual persistence and innovation variability. Some temporal regimes are geographically extensive and highly stable, whereas others have uncertain boundaries or are represented by small groups. Figure~\ref{fig:vi_partition} summarises this structure through the VI-optimal partition and its mean within-cluster co-clustering probability. The latter is essential for interpretation: the representative partition provides a convenient spatial summary, but boundaries involving low-stability groups should not be interpreted as sharp environmental discontinuities.

The posterior mean number of occupied clusters was 107.6, with median 108 and a 95\% posterior interval of $[102,113]$. The VI-optimal representative partition contained 119 clusters, including 20 singletons. Cluster sizes ranged from 1 to 485, with lower quartile 4, median 13, and upper quartile 25. The representative partition need not have the posterior mean or modal number of clusters because it minimises posterior expected variation-of-information loss rather than estimating $K$ directly \citep{wade2018,dahl2022}. Accordingly, these groups summarise posterior similarities in local temporal behaviour and should not be interpreted as 119 sharply delineated pollution regions.

The two regime parameters describe distinct aspects of local behaviour. Persistence $\phi$ measures the temporal dependence from one day to the next, whereas $\tau$ quantifies local innovation variability after removal of the shared spatial component. Regimes can therefore be interpreted along two complementary axes, ranging from relatively stable weakly persistent behaviour to persistent or highly variable local dynamics.

For the VI groups, posterior mean persistence ranges from 0.359 to 0.978, with median 0.681. Posterior mean local innovation standard deviations on the original PM10 scale range from 0.91 to 58.35, with median 5.41. Table~\ref{tab:largest_clusters} illustrates why cluster size and allocation stability should accompany these parameter summaries. The two largest groups contain 485 and 204 locations and have nearly unanimous within-group co-clustering, whereas several smaller representative groups have substantially lower stability. The latter should be interpreted as summaries of uncertain temporal-regime membership rather than equally well-defined classes.

\begin{table}[t]
\centering
\footnotesize
\caption{Eight largest clusters in the VI-optimal partition. Local innovation standard deviations $\tau$ are expressed in $\mu\mathrm{g}/\mathrm{m}^3$.}
\label{tab:largest_clusters}
\begin{tabular}{rrrrr}
\toprule
VI label & Size & Mean $\phi$ & Mean $\tau$ & Mean co-clustering\\
\midrule
6  & 485 & 0.698 & 0.91 & 0.999\\
8  & 204 & 0.743 & 1.29 & 1.000\\
7  & 110 & 0.681 & 3.37 & 0.479\\
10 & 84  & 0.648 & 2.09 & 0.633\\
9  & 84  & 0.663 & 2.81 & 0.485\\
14 & 79  & 0.693 & 1.86 & 0.733\\
2  & 74  & 0.683 & 2.45 & 0.650\\
19 & 69  & 0.629 & 2.42 & 0.679\\
\bottomrule
\end{tabular}
\end{table}

The largest stable group includes monitoring locations in Germany, France, Belgium and Austria, illustrating that similar residual dynamics can extend across national boundaries. The two largest groups combine relatively low local innovation variability with different persistence levels. In contrast, the third-largest group, represented particularly in Italy, Poland and the Czech Republic, has substantially weaker allocation stability despite persistence comparable to that of the largest group. Thus, similar persistence does not by itself define a well-supported temporal regime: local variability and uncertainty in membership also matter. These patterns describe the fitted residual dynamics, not shared emissions sources or contiguous pollution zones.

\begin{figure}[t]
\centering
\begin{overpic}[width=\textwidth]{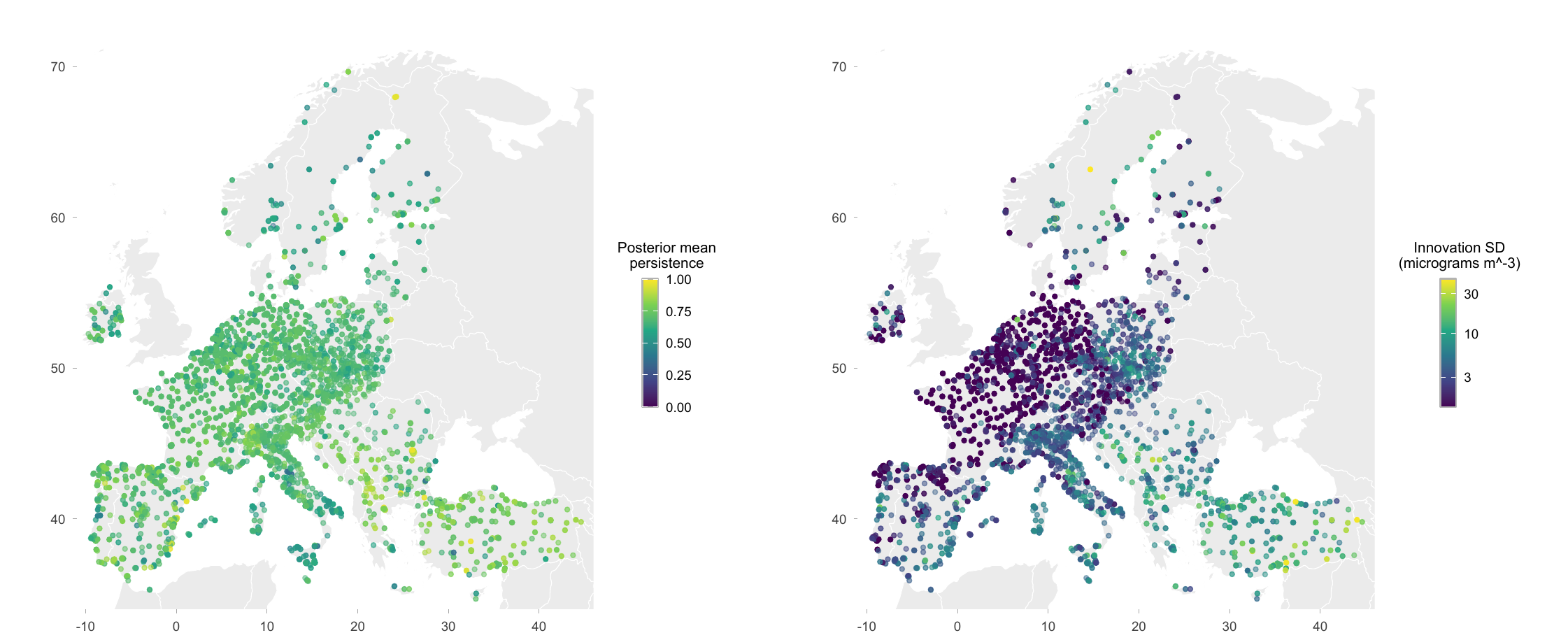}
\put(0.75,39.3){\jrsscpanellabel{A}}
\put(51.3,39.3){\jrsscpanellabel{B}}
\end{overpic}
\caption{Temporal-regime parameters summarised within VI groups: (A) posterior mean persistence $\phi$; (B) posterior mean local innovation standard deviation $\tau$, expressed in $\mu\mathrm{g}/\mathrm{m}^3$.}
\label{fig:regime_parameters}
\end{figure}

Figure~\ref{fig:regime_parameters} maps posterior mean persistence and local innovation standard deviation summarised within the VI groups. These quantities distinguish how strongly residual departures propagate between successive days from how much unpredictable local variation remains after accounting for the common shock. Groups with similar persistence can nevertheless differ substantially in innovation variability. The innovation standard deviation $\tau_k$ is shown on the original PM10 scale, rather than its square, to facilitate interpretation across groups with markedly different variability.

\subsubsection{Temporal regimes at unmonitored locations}
\label{sec:pm10_predictive_regimes}

The predictive partition extends the local temporal-regime structure beyond the observed monitoring network. For each unmonitored grid location, the posterior predictive distribution yields both regime-specific parameter summaries and a measure of allocation uncertainty. Because the posterior partition contains a large number of groups and cluster labels are not directly comparable across MCMC iterations, we focus on label-invariant predictive summaries of the local temporal characteristics. In particular, we report the posterior predictive means of persistence and innovation variability, together with a summary of predictive regime certainty.

\begin{figure}[t]
\centering
\begin{overpic}[width=\textwidth]{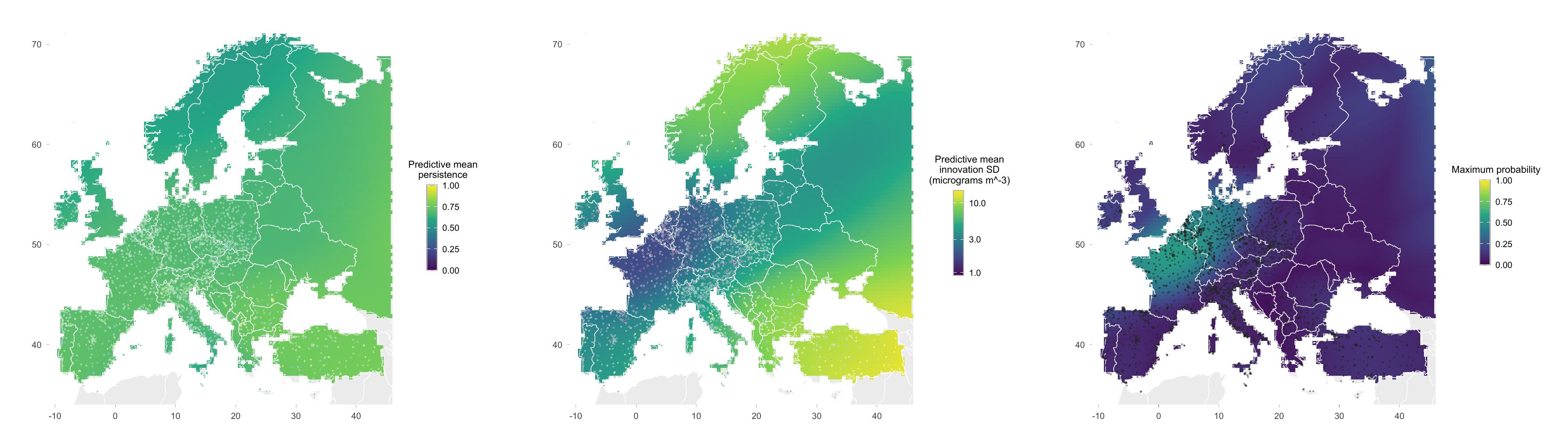}
\put(0.5,26.1){\jrsscpanellabel{A}}
\put(34.1,26.1){\jrsscpanellabel{B}}
\put(67.65,26.1){\jrsscpanellabel{C}}
\end{overpic}
\caption{Predictive summaries at unmonitored land locations: (A) mean persistence; (B) mean local innovation standard deviation in $\mu\mathrm{g}/\mathrm{m}^3$; (C) the largest VI-projected predictive regime probability. Panel C measures allocation certainty, not response uncertainty.}
\label{fig:predictive_temporal_regime}
\end{figure}

Figure~\ref{fig:predictive_temporal_regime} shows that the temporal-regime structure inferred at monitoring sites extends nontrivially across the wider domain. The predictive persistence surface indicates how strongly residual departures are expected to propagate from one day to the next, whereas the innovation panel summarises the expected magnitude of location-specific short-term variability after accounting for the shared spatial component. Together, these panels provide a spatially resolved description of local temporal behaviour that complements the VI partition at the observed stations.

Across the 5,297 prediction locations, posterior predictive mean persistence ranges from 0.547 to 0.780, with an interquartile range of $[0.652,0.708]$. Predictive mean local innovation standard deviation on the original PM10 scale ranges from 1.73 to 13.49, with an interquartile range of $[3.96,7.78]$. The predictive probability of a new regime remains small, averaging 0.001 and reaching at most 0.0094; it is not the dominant predictive category at any grid location. Thus, prediction primarily extends temporal regimes represented in the monitored network, while allowing uncertainty about which regime applies at each unmonitored location.

Panel (C) reports the largest predictive probability associated with a VI reference group or the new-regime category, calculated as described in Supplementary Section~\ref{supp:prediction}. Higher values indicate that probability is more concentrated on one category, whereas lower values indicate that several alternatives remain plausible. At an unmonitored location, regime membership is inferred rather than observed, so the dominant category should be interpreted together with its predictive probability.

A categorical dominant-regime map provides the corresponding discrete representation of the predictive partition and is presented in Supplementary Section~\ref{supp:pm10_predictive_regimes}. Because the representative partition contains more than one hundred groups, however, the categorical map is less effective for conveying broad spatial patterns than the label-invariant predictive summaries in Figure~\ref{fig:predictive_temporal_regime}. We therefore use the predictive $(\phi,\tau)$ surfaces and regime-certainty map as the primary representation of temporal behaviour between monitoring sites.

\subsubsection{Continuous prediction across the monitoring domain}

The fitted surfaces correspond to distinct environmental inferential targets. The common shock $u_t(\bs{s})$ represents spatially shared residual fluctuations and therefore describes the broad continuous spatial component of the model. The latent state $w_t(\bs{s})$ additionally reflects local persistence and regime-specific innovation, while the response $y_t(\bs{s})$ adds the measured meteorological contribution and observation-level variation. Examining these surfaces separately therefore reveals how broad spatial structure and local temporal behaviour combine to produce the observed PM10 field.

\begin{figure}[t]
\centering
\begin{overpic}[width=\textwidth]{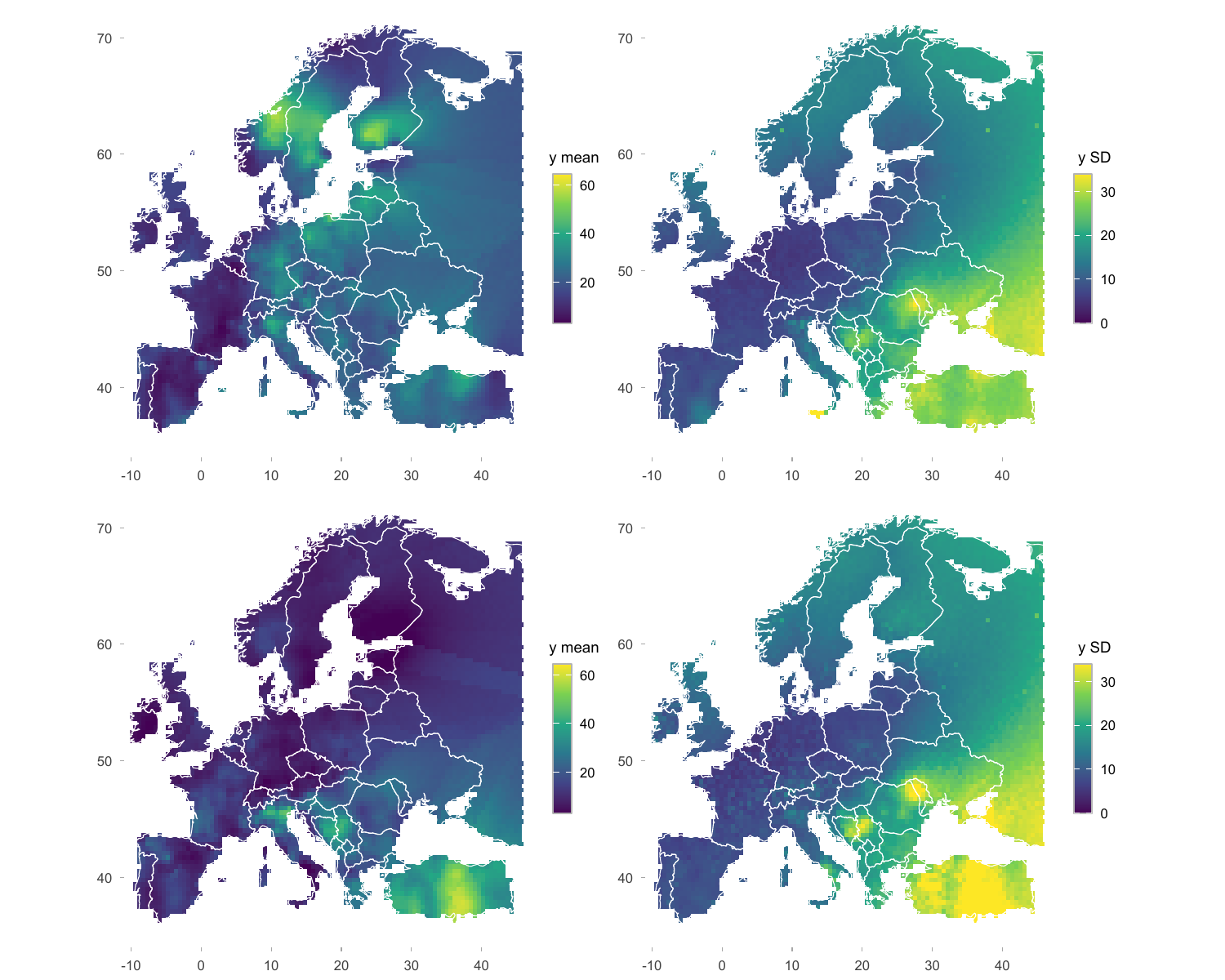}
\put(1.4,77.9){\jrsscpanellabel{A}}
\put(51.4,77.9){\jrsscpanellabel{B}}
\put(1.4,37.9){\jrsscpanellabel{C}}
\put(51.4,37.9){\jrsscpanellabel{D}}
\end{overpic}
\caption{Posterior PM10 response prediction: predictive mean (A, C) and predictive standard deviation (B, D) for April 9 (A, B) and December 15 (C, D) 2024. All quantities are expressed in $\mu\mathrm{g}/\mathrm{m}^3$. Corresponding quantities use common colour scales across dates.}
\label{fig:response_surfaces}
\end{figure}

Online prediction retained posterior means and standard deviations across the land grid and all 366 days. Figure~\ref{fig:response_surfaces} presents the response mean and standard deviation for two representative days, using common colour scales for corresponding quantities across dates. Spatially varying uncertainty reflects uncertainty in the common shock and temporal regime, together with local innovation and measurement variation. The corresponding Gaussian-derived 95\% predictive limits are presented in Supplementary Section~\ref{supp:response_prediction_maps}. These pointwise limits are Gaussian approximations based on predictive moments, not empirical posterior quantiles or simultaneous uncertainty bands.

The two dates illustrate how concentration mapping and temporal-regime mapping provide complementary information. On April 9, elevated predicted concentrations occur in parts of Scandinavia and central Europe, whereas the December 15 surface places greater concentrations in parts of southern and south-eastern Europe, including northern Italy and Turkey. Predictive standard deviations show a more persistent geographical pattern across the two dates. Thus, changes in the predicted concentration field need not coincide with changes in the underlying temporal-regime description, and similar predicted concentrations need not imply similar predictive uncertainty.

\subsubsection{Global parameters and variance decomposition}

Table~\ref{tab:application_parameters} reports posterior summaries of the global coefficients and spatial parameters on the standardised scale. Temperature has a positive conditional association with daily PM10, whereas precipitation and wind speed have negative associations. These coefficients adjust the temporal-regime and spatial-process summaries for measured meteorological variation; they describe conditional associations rather than causal effects. The estimated variance components further distinguish observation-level variation from the spatially shared fluctuations represented by the common shock.

\begin{table}[t]
\centering
\footnotesize
\caption{Posterior summaries for global and spatial parameters in the European PM10 analysis. Regression coefficients and variance parameters refer to the standardised modelling scale; the spatial range $\kappa$ is expressed in units of 1,000 km.}
\label{tab:application_parameters}
\begin{tabular}{lrrrr}
\toprule
Parameter & Mean & SD & 2.5\% & 97.5\%\\
\midrule
Intercept & 0.02125 & 0.00434 & 0.01224 & 0.02978\\
Temperature & 0.10870 & 0.00287 & 0.10338 & 0.11463\\
Precipitation & $-0.01388$ & 0.00076 & $-0.01538$ & $-0.01242$\\
Wind speed & $-0.05615$ & 0.00104 & $-0.05813$ & $-0.05417$\\
$\sigma^2$ & 0.004864 & 0.000052 & 0.004765 & 0.004970\\
$\sigma_u^2$ & 0.23897 & 0.00250 & 0.23438 & 0.24405\\
$\kappa$ & 0.43709 & 0.00510 & 0.42730 & 0.44712\\
\bottomrule
\end{tabular}
\end{table}

Overall acceptance rates were 0.440 for persistence updates and 0.226 for the spatial update. Effective sample sizes and block-specific timings are reported in Supplementary Section~\ref{supp:application_diagnostics}.

The regression coefficients exhibit substantial autocorrelation in the retained sample. Their summaries should therefore be interpreted with explicit acknowledgement of their limited effective sample sizes; the within-chain diagnostics do not establish convergence of the joint posterior.

\begin{figure}[t]
\centering
\begin{overpic}[width=\textwidth]{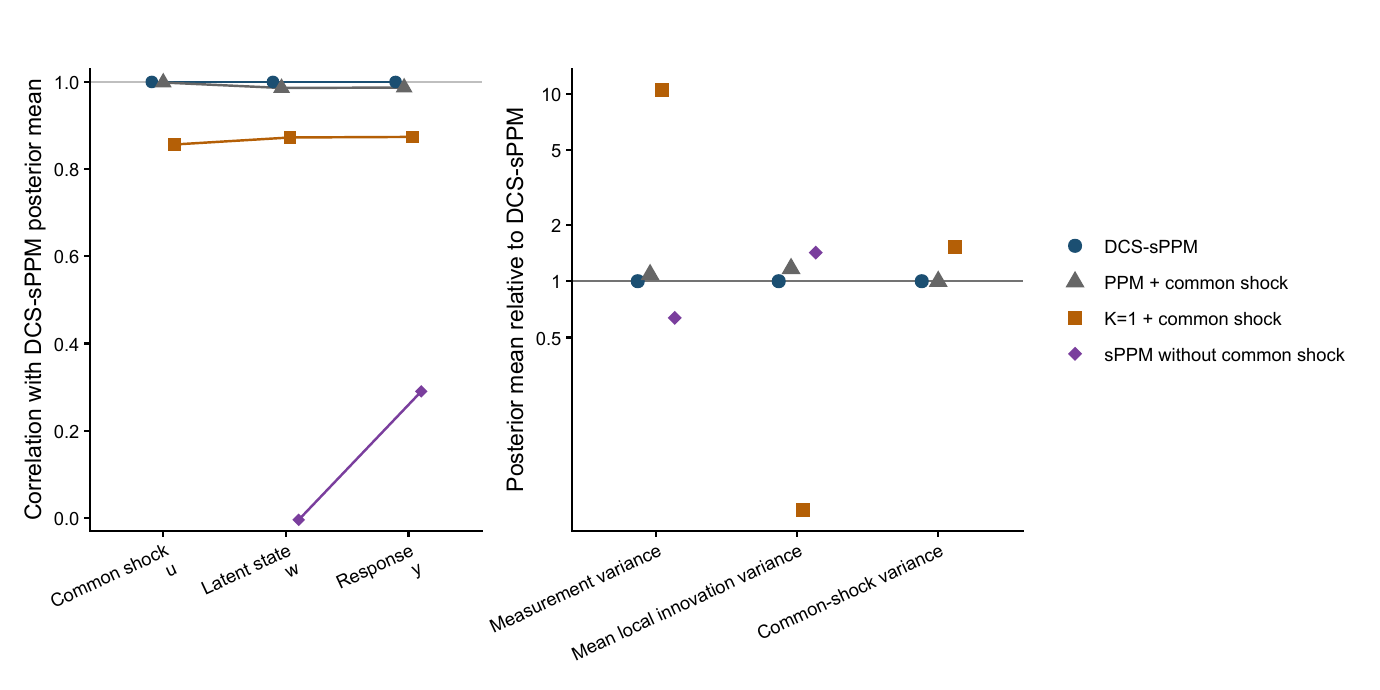}
\put(1.5,48.6){\jrsscpanellabel{A}}
\put(37,48.6){\jrsscpanellabel{B}}
\end{overpic}
\caption{Structural sensitivity of the European PM10 analysis across all four models. (A) Correlations of posterior mean prediction surfaces with DCS-sPPM; the no-shock model has no common-shock surface by construction. (B) Posterior mean variance components relative to DCS-sPPM, shown on a logarithmic scale.}
\label{fig:comparative_application}
\end{figure}

The four-model comparison clarifies the empirical roles of the proposed components (Figure~\ref{fig:comparative_application}). Removing spatial cohesion preserves a similar posterior mean response surface, with correlation 0.987 relative to the DCS-sPPM, but produces more geographically dispersed representative groups and approximately 9\% higher average predictive standard deviation. Imposing a single temporal regime changes the response surface more substantially and increases the posterior mean measurement-error variance $\sigma^2$ by a factor of approximately 10. Removing the common shock yields the strongest change in the spatial response pattern and approximately 20\% higher average predictive standard deviation. These are structural sensitivity comparisons rather than out-of-sample accuracy assessments: their contribution is to show how geographic regime information, heterogeneous local dynamics, and continuous spatial variation affect distinct aspects of inference. Complete results are reported in Supplementary Section~\ref{supp:comparative_application}.

\section{Discussion}
\label{sec:discussion}

Environmental monitoring networks can contain two forms of structure that are easy to conflate: smoothly shared spatial variation and heterogeneity in the local dynamics through which departures persist over time. The results of this study show that separating these components can provide a useful description of both. The continuous common-shock process captures residual variation shared across neighbouring locations, while the spatial partition identifies groups characterised by similar local persistence and innovation variability. This distinction also extends naturally to prediction, where spatial interpolation and temporal-regime assignment represent complementary sources of information at unmonitored locations.

The simulation study clarifies when this decomposition is informative. When temporal regimes are well separated, the local time-series information is sufficiently strong that the partition can be recovered even when the labels have little geographic organisation. When regimes are weakly separated, spatial coherence becomes considerably more valuable. Conversely, when true regime membership is unrelated to location, coordinates alone cannot predict the regime of an unobserved site. The spatial prior should therefore be understood as a scientifically motivated source of information, not as a generic device for resolving poorly separated mixtures.

A second finding is that uncertainty about temporal regimes does not imply failure of continuous environmental prediction. The common spatial component can remain accurately predicted even when the discrete regime at a new location is uncertain, and posterior averaging propagates that regime uncertainty into prediction of the latent state and observed response. This distinction is particularly relevant for monitoring applications, where characterising local dynamics and reconstructing the environmental field are related but not identical inferential goals.

A central feature of the DCS-sPPM is that clustering and spatial prediction are performed within the same probabilistic framework. The partition identifies groups of locations sharing similar local temporal dynamics, while the common-shock process provides a continuously varying spatial representation that can be interpolated at unmonitored sites. Prediction follows directly from the joint posterior rather than from separately fitted clustering and prediction models, accounting for uncertainty in regime allocation, spatial interpolation, local temporal evolution, and measurement error. This is particularly relevant for environmental monitoring networks, where interest often lies both in identifying locations with similar temporal behaviour and in predicting the environmental process between observed stations.

The European analysis shows that similar persistence need not imply similar local variability or equally reliable regime membership. The largest stable groups span several countries, whereas other groups with comparable persistence have substantially weaker allocation stability. Thus, neither national boundaries nor persistence alone provides the temporal-regime description delivered by the fitted model. The structural comparisons further separate the inferential roles of its components: removing geographic similarity changes regime organisation and predictive uncertainty while preserving a similar mean response surface; removing the common shock changes the spatial response pattern much more strongly. These comparisons do not establish out-of-sample superiority, but show why concentration mapping and residual temporal-regime mapping answer different questions within the same monitoring network.

The present formulation assumes a stationary spatial correlation shared across time and restricts local persistence to positive values. Predictive regime allocation uses geographic information when no response trajectory is available at the new location, making its informativeness dependent on the spatial organisation of temporal regimes. The meteorological coefficients describe conditional associations rather than causal effects. The full-data comparisons assess structural sensitivity, complementing the replicated prediction experiments rather than constituting held-out real-data accuracy assessments.

Several modelling extensions are natural. More flexible local dynamics could accommodate negative persistence by replacing the current Beta base measure with a suitable transformation onto $(-1,1)$. Dynamic or nonstationary common shocks could allow the spatial dependence structure to evolve over time and space \citep{higdon2022nonstationary,huang2025nonstationary,coube2026nonstationary,pereira2026spde}, while covariate-dependent similarity functions could incorporate additional geographic information such as topography or monitoring class. If direct interpretation of $\bs{\beta}$ becomes a primary inferential objective, restricted spatial regression, orthogonalisation, or stronger scientific priors could also be considered \citep{reich2006effects,hodges2010adding}; the present unconstrained decomposition was instead chosen to preserve predictive flexibility.

A particularly promising direction concerns neural approaches to reducing the computational burden of the model. Recent stochastic-process emulators learn generative approximations to Gaussian-process priors while retaining dependence on their covariance hyperparameters, allowing the learned representation to remain embedded within Bayesian inference \citep{semenova2022priorvae,navott2026deeprv}. Such constructions could provide an alternative to the current NNGP representation while leaving the temporal-regime partition and its interpretation unchanged. A complementary possibility is amortised Bayesian inference, in which simulations from the generative model are used to train a reusable neural posterior approximator \citep{papamakarios2016,radev2023bayesflow,maceda2024variational}. This could be especially useful for monitoring systems that are analysed repeatedly as new data arrive or alternative model specifications are considered. Fully amortising inference for the DCS-sPPM is nevertheless challenging because the partition is discrete and variable-dimensional, suggesting that a hybrid strategy may be more natural, with neural approximation used for the spatial process, continuous parameters, or selected predictive quantities while retaining explicit Bayesian updating for the partition.

The broader contribution of the DCS-sPPM is therefore not simply to combine a clustering model with a spatial predictor. It provides two complementary representations of an environmental monitoring network: a continuous spatial process describing variation shared across locations and a discrete, interpretable representation of heterogeneity in local temporal dynamics. Extending both representations beyond the observed network while propagating their uncertainty provides a framework for moving from site-specific time series toward probabilistic maps of environmental behaviour. The European PM10 analysis demonstrates that this can be done at continental scale while retaining uncertainty in both the latent environmental surface and the local temporal regimes.

\section*{Data and code availability}

PM10 observations and monitoring metadata were obtained from the European Environment Agency Air Quality Download Service \citep{eea_air_quality_2025}, and meteorological covariates were obtained from the Copernicus Climate Change Service ERA5 data product \citep{c3s_era5_single_levels}. The reproduction code is available at \url{https://github.com/lucaaiello/DCS-sPPM}. The repository README provides access to the versioned release containing the data inputs used in the European analysis, together with input descriptions, and installation instructions. The repository also documents software requirements and commands for reproducing the simulation experiments, model fits, numerical summaries and figures.

\setcounter{section}{0}
\setcounter{subsection}{0}
\setcounter{equation}{0}
\setcounter{table}{0}
\setcounter{figure}{0}

\renewcommand{\thesection}{S\arabic{section}}
\renewcommand{\thesubsection}{S\arabic{section}.\arabic{subsection}}
\renewcommand{\theequation}{S\arabic{equation}}
\renewcommand{\thetable}{S\arabic{table}}
\renewcommand{\thefigure}{S\arabic{figure}}

\section{Posterior computation and implementation}
\label{supp:posterior}

This section provides the posterior-computation details underlying the
sampler summarised in Section~\ref{sec:priors_inference} of the main text.
We first describe the principal sampling blocks and then collect the
implementation details used by the kernel. Constants independent
of the updated quantity are omitted throughout.
The derivations below concern the implemented exponential correlation specified in Section~\ref{sec:priors_inference}, for which the correlation parameters reduce to the scalar range $\kappa$.

Let
$\bs{Y}=(\bs{y}_1,\ldots,\bs{y}_n)^\top$ denote the matrix of sitewise
trajectories, and let $\bs{X}$ denote the design matrix stacked in
time-major order. Given allocations $\bs{c}$ and occupied-regime parameters
$\{\phi_k^*,\tau_k^{2*}\}_{k=1}^K$, write
$\phi_i=\phi_{c_i}^*$ and $\tau_i^2=\tau_{c_i}^{2*}$.
The joint posterior is proportional to
\begin{align*}
 &p(\bs{y}\mid\bs{\beta},\bs{w},\sigma^2)
 \prod_{i=1}^n
 p(\bs{w}_i\mid\bs{u}_i,\phi_i,\tau_i^2)
 \prod_{t=2}^T
 p(\bs{u}_t\mid\kappa,\sigma_u^2) \\
 &\qquad\times
 p(\rho_n\mid\bs{s})
 \prod_{k=1}^K
 p(\phi_k^*)p(\tau_k^{2*})
 p(\bs{\beta}\mid\bs{\Xi})
 p(\bs{\Xi})
 p(\sigma^2)
 p(\kappa)
 p(\sigma_u^2).
\end{align*}

For later use, define the site-specific transition sum of squares
\begin{equation}
\label{eq:supp_A_i}
 A_i(\phi)
 =
 (1-\phi^2)w_{i1}^2
 +
 \sum_{t=2}^T
 \{w_{it}-\phi w_{i,t-1}-u_{it}\}^2.
\end{equation}
The corresponding state density satisfies
\begin{equation}
\label{eq:supp_site_state_density}
 p(\bs{w}_i\mid\bs{u}_i,\phi,\tau^2)
 \propto
 (1-\phi^2)^{1/2}
 (\tau^2)^{-T/2}
 \exp\{-A_i(\phi)/(2\tau^2)\}.
\end{equation}
The stationary initialisation contributes the first term in
equation~\eqref{eq:supp_A_i} and accounts for the power $T/2$ in
equation~\eqref{eq:supp_site_state_density}.

\subsection{Posterior sampling}
\label{supp:posterior_sampling}

\paragraph{Collapsed update of the global regression coefficients.}

For site $i$, let $\bs{L}_i$ be the lower-bidiagonal autoregressive operator with diagonal $(\sqrt{1-\phi_i^2},1,\ldots,1)$ and subdiagonal entries $-\phi_i$, and define $\bs{u}_i^*=(0,u_2(\bs{s}_i),\ldots,u_T(\bs{s}_i))^\top$. Then
\[
  \bs{Q}_i
  =
  \tau_i^{-2}\bs{L}_i^\top\bs{L}_i,
  \qquad
  \bs{h}_i
  =
  \tau_i^{-2}\bs{L}_i^\top\bs{u}_i^*,
\]
so that
\[
  \bs{w}_i
  \mid
  \bs{u}_i,\phi_i,\tau_i^2
  \sim
  \mathcal{N}_T
  \left(
    \bs{Q}_i^{-1}\bs{h}_i,
    \bs{Q}_i^{-1}
  \right).
\]
Integrating $\bs{w}_i$ out of the observation model gives
\[
  \bs{y}_i
  \mid
  \bs{\beta},\text{rest}
  \sim
  \mathcal{N}_T
  \left(
    \bs{X}_i\bs{\beta}+\bs{Q}_i^{-1}\bs{h}_i,
    \bs{Q}_i^{-1}+\sigma^2\bs{I}_T
  \right).
\]
Let
\[
  \bs{S}_i=\bs{Q}_i^{-1}+\sigma^2\bs{I}_T.
\]
Combining the sitewise contributions with the Gaussian prior yields
\[
  \bs{\beta}\mid\text{rest}
  \sim
  \mathcal{N}_p
  \left(
    \widetilde{\bs{V}}_\beta
    \widetilde{\bs{b}}_\beta,
    \widetilde{\bs{V}}_\beta
  \right),
\]
where
\begin{align*}
  \widetilde{\bs{V}}_\beta^{-1}
  &=
  \bs{\Xi}^{-1}
  +
  \sum_{i=1}^n
  \bs{X}_i^\top
  \bs{S}_i^{-1}
  \bs{X}_i,\\
  \widetilde{\bs{b}}_\beta
  &=
  \sum_{i=1}^n
  \bs{X}_i^\top
  \bs{S}_i^{-1}
  \left(
    \bs{y}_i-\bs{Q}_i^{-1}\bs{h}_i
  \right).
\end{align*}
The Woodbury identity reduces multiplication by $\bs{S}_i^{-1}$ to
tridiagonal solves with
$\bs{Q}_i+\sigma^{-2}\bs{I}_T$; neither $\bs{S}_i$ nor its inverse is
formed explicitly.

\paragraph{Sitewise latent trajectories.}

For $T>1$, the nonzero entries of the autoregressive precision
$\bs{Q}_i$ are
\begin{align*}
  &[\bs{Q}_i]_{11}
  =
  \tau_i^{-2},\\
  &[\bs{Q}_i]_{TT}
  =
  \tau_i^{-2},\\
  &[\bs{Q}_i]_{tt} 
  =
  (1+\phi_i^2)\tau_i^{-2}, & t&=2,\ldots,T-1 \\
  &[\bs{Q}_i]_{t,t+1} 
  =
  [\bs{Q}_i]_{t+1,t}
  =
  -\phi_i\tau_i^{-2} & t&=1,\ldots,T-1,
\end{align*}
The contribution of the common shock to the canonical vector is
\begin{align*}
  h_{i,t}
  &=
  \tau_i^{-2}
  \{u_t(\bs{s}_i)-\phi_i u_{t+1}(\bs{s}_i)\},
  &&
  t=1,\ldots,T-1,\\
  h_{i,T}
  &=
  \tau_i^{-2}u_T(\bs{s}_i),
\end{align*}
using the convention $u_1(\bs{s}_i)=0$.

After adding the observation contribution,
\[
  \bs{w}_i\mid\text{rest}
  \sim
  \mathcal{N}_T
  \left(
    \bs{P}_{w_i}^{-1}\bs{b}_{w_i},
    \bs{P}_{w_i}^{-1}
  \right),
  \qquad
  \bs{P}_{w_i}
  =
  \bs{Q}_i+\sigma^{-2}\bs{I}_T.
\]
where $\bs{b}_{w_i}=\sigma^{-2}(\bs{y}_i-\bs{X}_i\bs{\beta})+\bs{h}_i$.
Because $\bs{P}_{w_i}$ is tridiagonal, factorisation and sampling require
$O(T)$ operations and $O(T)$ working memory for each site.

\paragraph{Timewise common shocks.}

For $t=2,\ldots,T$, define
\[
  \bs{a}_t
  =
  \bs{w}_t-\bs{\Phi}\bs{w}_{t-1}.
\]
Under the state equation,
\[
  \bs{a}_t
  =
  \bs{u}_t+\bs{\eta}_t,
  \qquad
  \bs{\eta}_t
  \sim
  \mathcal{N}_n(\bs{0},\bs{\mathcal T}),
\]
where $\bs{\Phi}=\diag(\phi_1,\ldots,\phi_n)$
and $\bs{\mathcal T}=\diag(\tau_1^2,\ldots,\tau_n^2)$.
Combining this likelihood with the NNGP prior gives
\[
  \bs{u}_t\mid\text{rest}
  \sim
  \mathcal{N}_n
  \left(
    \bs{P}_u^{-1}\bs{b}_{u,t},
    \bs{P}_u^{-1}
  \right),
\]
with
\[
  \bs{P}_u
  =
  \bs{Q}_u(\kappa,\sigma_u^2)
  +
  \bs{\mathcal T}^{-1},
  \qquad
  \bs{b}_{u,t}
  =
  \bs{\mathcal T}^{-1}\bs{a}_t.
\]
All $T-1$ time points share the same precision matrix within an MCMC iteration. The implementation therefore performs one sparse numerical Cholesky factorisation of $\bs{P}_u$ and reuses it for all timewise solves and draws. Because the neighbour graph fixes its sparsity pattern, the symbolic analysis and fill-reducing permutation are computed once and reused throughout sampling.

\paragraph{Measurement variance and regression-prior scales.}

Let
\[
  \mathrm{SSE}
  =
  \sum_{i=1}^n\sum_{t=1}^T
  \{
    y_t(\bs{s}_i)
    -
    \bs{x}_t(\bs{s}_i)^\top\bs{\beta}
    -
    w_t(\bs{s}_i)
  \}^2.
\]
The corresponding full conditional distributions are
\begin{align*}
  \sigma^2\mid\text{rest}
  &\sim
  \mathrm{IG}
  \left(
    a_\sigma+\frac{nT}{2},
    b_\sigma+\frac{\mathrm{SSE}}{2}
  \right),\\
  \xi_l^2\mid\text{rest}
  &\sim
  \mathrm{IG}
  \left(
    a_\xi+\frac12,
    b_\xi+\frac{\beta_l^2}{2}
  \right),
  \qquad
  l=1,\ldots,p.
\end{align*}
The regression-prior scales are sampled from their inverse-Gamma full conditionals. The implemented variance update uses joint Metropolis--Hastings moves for $\sigma^2$ and each occupied regime variance $\tau_k^{2*}$, integrating out the latent trajectories. Proposals are made on $(\log\sigma^2,\log(\tau_k^{2*}/\sigma^2))^\top$, with the corresponding Jacobian included. The trajectories are subsequently regenerated; following the allocation and persistence updates, the regime variances are additionally sampled from their inverse-Gamma full conditionals.

\paragraph{Allocation update.}

After removing site $i$ from its current group, the log weight for assigning
it to an occupied cluster $S_k^{(-i)}$ is
\begin{equation}
\label{eq:supp_existing_weight}
 \log\omega_{ik}
 =
 \ell_i(\phi_k^*,\tau_k^{2*})
 +
 \log
 \frac{C(S_k^{(-i)}\cup\{i\})}
      {C(S_k^{(-i)})}
 +
 \log
 \frac{
   g(\bs{s}_k^{*(-i)}\cup\{\bs{s}_i\})
 }{
   g(\bs{s}_k^{*(-i)})
 },
\end{equation}
where
\[
  \ell_i(\phi,\tau^2)
  =
  \frac12\log(1-\phi^2)
  -
  \frac{T}{2}\log\tau^2
  -
  \frac{A_i(\phi)}{2\tau^2}.
\]

New-cluster proposals are generated from the base measure. For an auxiliary
candidate
$(\phi_h^{\mathrm{aux}},\tau_h^{2,\mathrm{aux}})$,
\begin{equation}
\label{eq:supp_aux_weight}
 \log\omega_{ih}^{\mathrm{aux}}
 =
 \ell_i
 \left(
   \phi_h^{\mathrm{aux}},
   \tau_h^{2,\mathrm{aux}}
 \right)
 +
 \log C(\{i\})
 +
 \log g(\{\bs{s}_i\})
 -
 \log K_{\mathrm{aux}}.
\end{equation}
When removal of site $i$ destroys a singleton, its former parameter value is included among the auxiliary candidates. The weights in equations~\eqref{eq:supp_existing_weight}--\eqref{eq:supp_aux_weight} are normalised on the logarithmic scale to ensure numerical stability before categorical sampling.

For the Dirichlet process (DP) cohesion used in the paper, the occupied-cluster
cohesion ratio is $|S_k^{(-i)}|$, while the singleton cohesion contribution is \(M\) \citep{neal2000markov}.

To avoid rescanning complete trajectories for every candidate regime, the
implementation stores the coefficients of
\[
  A_i(\phi)
  =
  a_{i0}-2a_{i1}\phi+a_{i2}\phi^2.
\]
The implementation also caches cluster coordinate counts, sums and cross-products required by the spatial similarity function. These statistics are updated when a site is removed from its current cluster and when it is assigned to its sampled destination.

\paragraph{Regime-specific temporal parameters.}

For cluster $S_k$, define
\[
  A_k(\phi)
  =
  \sum_{i\in S_k}A_i(\phi).
\]
Conditional on $\tau_k^{2*}$, the persistence parameter has log target
\[
  \log\pi(\phi_k^*\mid\text{rest})
  =
  (a_\phi-1)\log\phi_k^*
  +
  (b_\phi-1)\log(1-\phi_k^*)
  +
  \sum_{i\in S_k}
  \ell_i(\phi_k^*,\tau_k^{2*}).
\]
A random-walk Metropolis proposal is applied on the logit scale and includes
the associated transformation Jacobian. Given the current partition, latent trajectories, common shocks, and persistence parameter $\phi_k^*$, the innovation variance has full conditional distribution
\[
  \tau_k^{2*}\mid\text{rest}
  \sim
  \mathrm{IG}
  \left(
    a_\tau+\frac{|S_k|T}{2},
    b_\tau+\frac{A_k(\phi_k^*)}{2}
  \right).
\]

\paragraph{Collapsed update of the spatial parameters.}

The NNGP likelihood can be written without forming a dense covariance matrix.
Using the NNGP innovation representation,
\begin{align*}
  \log p(\bs{u}_{2:T}\mid\kappa,\sigma_u^2)
  ={}&
  -\frac{n(T-1)}{2}\log\sigma_u^2
  -\frac{T-1}{2}
   \sum_{i=1}^n\log\delta_i(\kappa)\\
  &-
  \frac{1}{2\sigma_u^2}
  \sum_{t=2}^T
  \left\|
    \bs{\Delta}(\kappa)^{-1/2}
    \bs{B}(\kappa)\bs{u}_t
  \right\|^2.
\end{align*}

The kernel instead updates $(\kappa,\sigma_u^2)$ jointly after
integrating out $\bs{u}_{2:T}$ from the Gaussian transition likelihood.
Because
\[
  \bs{a}_t
  \mid
  \kappa,\sigma_u^2,\bs{\mathcal T}
  \sim
  \mathcal{N}_n
  \left(
    \bs{0},
    \bs{Q}_u^{-1}+\bs{\mathcal T}
  \right),
\]
let
\[
  \bs{P}_u
  =
  \bs{Q}_u+\bs{\mathcal T}^{-1},
  \qquad
  \bs{b}_{u,t}
  =
  \bs{\mathcal T}^{-1}\bs{a}_t.
\]
The determinant lemma and Woodbury identity give
\begin{align*}
  \log|\bs{Q}_u^{-1}+\bs{\mathcal T}|
  &=
  \log|\bs{P}_u|
  +
  \log|\bs{\mathcal T}|
  -
  \log|\bs{Q}_u|,\\
  \bs{a}_t^\top
  (\bs{Q}_u^{-1}+\bs{\mathcal T})^{-1}
  \bs{a}_t
  &=
  \bs{a}_t^\top
  \bs{\mathcal T}^{-1}
  \bs{a}_t
  -
  \bs{b}_{u,t}^\top
  \bs{P}_u^{-1}
  \bs{b}_{u,t}.
\end{align*}
Combining the integrated Gaussian likelihood with the priors gives the following log conditional posterior, where $\mathrm{const.}$ collects terms that do not depend on $(\kappa,\sigma_u^2)$:
\begin{align}
\label{eq:supp_collapsed_spatial}
 \log\pi(\kappa,\sigma_u^2\mid\text{rest})
 ={}&
 \log p(\kappa)
 +
 \log p(\sigma_u^2) \nonumber\\
 &-
 \frac{T-1}{2}
 \{
   \log|\bs{P}_u|
   -
   \log|\bs{Q}_u|
 \}
 +
 \frac12
 \sum_{t=2}^T
 \bs{b}_{u,t}^\top
 \bs{P}_u^{-1}
   \bs{b}_{u,t}
  +\mathrm{const}.
\end{align}
Gaussian random-walk proposals are applied to
\[
  \bs{\vartheta}
  =
  (\log\kappa,\log\sigma_u^2)^\top,
\]
with the corresponding Jacobian added to
equation~\eqref{eq:supp_collapsed_spatial}. Proposal covariance is adapted
during warm-up and then frozen.

After the accepted values of $(\kappa,\sigma_u^2)$ are fixed,
$\bs{u}_{2:T}$ is regenerated from its Gaussian conditional before any later
block conditions on it. The resulting procedure is therefore a partially
collapsed Gibbs/Metropolis--Hastings composition with the desired posterior
as invariant distribution \citep{van2008partially}.

\subsection{Implementation}
\label{supp:implementation}

Each iteration begins with a collapsed update of $\bs{\beta}$, followed by joint collapsed moves for the measurement and regime-specific innovation variances. The sitewise latent trajectories are then regenerated and the regression-prior scales updated. Conditional on these trajectories, the sampler updates allocations, persistence parameters and regime innovation variances. A joint collapsed spatial-parameter move is followed by parallel draws of the common shocks, after which the retained parameter and predictive summaries are updated.

The state vector $\bs{w}$ is therefore sampled only once per iteration.
This ordering does not reuse a state generated under an incompatible target:
the trajectory update precedes all blocks that condition on the refreshed
state, and $\bs{w}$ is updated again at the beginning of the following
iteration after the intervening parameter updates.

\paragraph{Computational complexity.}

For fixed neighbour size $m$, the NNGP graph contains $O(nm)$ directed
neighbour entries. The sitewise state updates and collapsed regression
calculations require $O(nT)$ work because their temporal systems are
tridiagonal. Construction of temporal sufficient statistics is also
$O(nT)$, after which candidate allocation likelihoods can be evaluated from
cached quadratic coefficients.

The common-shock block requires one sparse factorisation of an
$n\times n$ precision matrix and $T-1$ pairs of triangular solves. Its
exact complexity depends on sparse fill-in under the stored ordering.
For $n_0$ prediction locations and $B_p$ predictive draws, evaluating the common-shock predictive means requires $O(B_pTn_0m)$ arithmetic operations, while storing online means and variances requires only $O(Tn_0)$ memory for each predicted process. Neighbour covariance factorisations and predictive regime allocation incur additional computational costs.

\paragraph{Numerical implementation.}

Allocation weights are normalised on the logarithmic scale, and Metropolis--Hastings acceptance probabilities are computed from log-density differences. Persistence parameters are represented on the logit scale, while variances and spatial ranges are represented on the log scale. NNGP conditional variances are checked for positivity and finiteness.

Max--min ordering, neighbour indices, graph storage, sparse symbolic analysis, and design-matrix cross-products are constructed before the MCMC loop. Acceptance rates and block-level timings are retained as diagnostics.

\section{Prediction and posterior-partition summaries}
\label{supp:prediction}

This section provides additional details for the predictive construction in
Section~\ref{sec:prediction} and for the partition summaries used throughout
the simulation study and European PM10 analysis.

\subsection{Prediction at an unmonitored location}
\label{supp:newsite_prediction}

For a new site $\bs{s}_0$ appended after the observed NNGP ordering, let
$N(0)$ denote its $m$ selected observed neighbours,
$\bs{r}_{N0}(\kappa)$ the column vector of correlations between those neighbours and the new site, with $\bs{r}_{0N}(\kappa)=\bs{r}_{N0}(\kappa)^\top$, and $\bs{R}_{NN}(\kappa)$ denote the $m\times m$ correlation matrix among these neighbours. Define
\begin{align*}
  \bs{b}_0(\kappa)
  &=
  \bs{R}_{NN}(\kappa)^{-1}
  \bs{r}_{N0}(\kappa),\\
  \delta_0(\kappa)
  &=
  1-
  \bs{r}_{0N}(\kappa)
  \bs{R}_{NN}(\kappa)^{-1}
  \bs{r}_{N0}(\kappa).
\end{align*}
Because the covariance matrix and cross-covariance vector share the factor $\sigma_u^2$, this factor cancels when calculating the interpolation weights $\bs{b}_0(\kappa)$. The conditional variance is instead $\sigma_u^2\delta_0(\kappa)$, where $\delta_0(\kappa)$ is the conditional variance under unit marginal variance. Thus, conditional on the common-shock values at the monitored sites,
\[
  \widetilde{u}_t(\bs{s}_0)
  \mid
  \bs{u}_t,\kappa,\sigma_u^2
  \sim
  \mathcal{N}
  \left(
    \bs{b}_0(\kappa)^\top
    \bs{u}_t\{N(0)\},
    \sigma_u^2\delta_0(\kappa)
  \right).
\]

For a fixed posterior draw and prediction location, the
$m\times m$ neighbour covariance block is factorised once and reused across
all times. After the predictive regime has been sampled, the local state
trajectory is generated through equation~\eqref{eq:w_prediction}, and the
response is obtained from equation~\eqref{eq:y_prediction}.

Conditional on the sampled regime and model parameters, the local recursion
is Gaussian. Marginally, however, the predictive distribution is a mixture
over posterior allocations, regime-specific parameters, common-shock draws,
and global parameters. Posterior averaging is therefore carried out after
sampling from the complete predictive hierarchy rather than by substituting
posterior mean parameter values.

\paragraph{Online predictive moments.}

For a fixed prediction location and time point, let $z^{(b)}$ denote the predictive draw of one of the processes $u$, $w$, or $y$ at retained iteration $b$. The sampler maintains the number of draws, their running mean $\bar z_b$, and the sum of squared deviations $S_b=\sum_{r=1}^{b}(z^{(r)}-\bar z_b)^2$. Starting from $\bar z_1=z^{(1)}$ and $S_1=0$, these quantities are updated for $b\geq2$ as
\begin{align*}
  d_b &= z^{(b)}-\bar z_{b-1},\\
  \bar z_b &= \bar z_{b-1}+\frac{d_b}{b},\\
  S_b &= S_{b-1}+d_b\{z^{(b)}-\bar z_b\}.
\end{align*}
After $B$ predictive draws, the posterior predictive variance is estimated by $S_B/(B-1)$. These updates are applied separately to each process, prediction location, and time point, without storing the individual draws.

For
$q\in\{0.025,0.975\}$,
the pointwise Gaussian-derived limits at each prediction location and time point are
\[
  \widehat z_q
  =
  \bar z_B
  +
  \Phi^{-1}(q)
  \sqrt{
    \frac{S_{B}}{B-1}
  }.
\]
These are moment-matched Gaussian limits rather than empirical posterior quantiles and are used for the European predictive maps. In the simulation study, coverage was evaluated using empirical equal-tailed intervals calculated from the retained predictive draws.

\subsection{Representative partitions and predictive regime summaries}
\label{supp:partition_summaries}

Let $\rho$ and $\rho'$ denote two partitions and let $H(\rho)$ denote the
entropy of the empirical cluster-size distribution. Their variation of
information is
\[
  \mathrm{VI}(\rho,\rho')
  =
  H(\rho)+H(\rho')-2I(\rho,\rho'),
\]
where $I(\rho,\rho')$ is the mutual information induced by the
cross-classification of observations.

For the European application, the retained partitions $\rho^{(1)},\ldots,\rho^{(B)}$ are summarised by the representative partition
\[
  \widehat\rho_{\mathrm{VI}}
  =
  \arg\min_{\rho\in\mathcal{A}}
  \frac{1}{B}
  \sum_{b=1}^B
  \mathrm{VI}
  \{
    \rho,\rho^{(b)}
  \},
\]
where $\mathcal{A}$ is the set of candidate partitions considered by the optimisation
\citep{meilua2007,wade2018,dahl2022}.
Because variation of information depends only on the induced partition, this summary 
is invariant to MCMC label switching.

The VI-optimal partition is a decision-theoretic posterior summary: it is the partition that minimises the average information discrepancy from the retained posterior partitions. A value of $\mathrm{VI}(\rho,\rho')=0$ means that the two partitions induce exactly the same grouping, up to arbitrary label permutations; larger values indicate disagreement in how sites are grouped through splits or merges.

The set $\mathcal{A}$ denotes the candidate partitions considered by the
optimisation. Enumerating all partitions of 2,989 sites is infeasible, so the
implementation uses the retained MCMC allocations to characterise the
posterior distribution and a stochastic search to identify a partition with
low posterior expected VI loss. Thus, the MCMC draws describe posterior uncertainty, whereas $\widehat\rho_{\mathrm{VI}}$ provides a single, label-invariant representative partition.

The posterior similarity matrix is
\[
  \widehat p_{ij}
  =
  \frac{1}{B}
  \sum_{b=1}^{B}
  \mathbb{I}\{c_i^{(b)}=c_j^{(b)}\}.
\]
The stability of a VI group is the average of $\widehat p_{ij}$ over pairs of distinct sites within that group. Higher values indicate that its sites are more consistently assigned together across posterior draws, irrespective of numerical cluster labels. Singleton stability is set to one by convention because there are no within-group pairs; this value does not indicate stronger posterior support for a singleton regime.

For the simulation performance summaries, a representative sampled partition is selected by minimising $\sum_{i,j}\{\mathbb{I}(c_i^{(b)}=c_j^{(b)})-\widehat p_{ij}\}^2$ over retained draws $b$. This provides a label-invariant summary based on posterior pairwise co-membership. VI is subsequently used, together with the ARI and pairwise measures, to evaluate recovery against the data-generating partition.

\paragraph{Predictive regime probabilities relative to the VI partition.}

To summarise the predicted temporal regime at an unmonitored location, we use the groups of the VI-optimal partition as a common reference. Individual MCMC draws may contain different numbers of clusters and different groupings of the observed sites; moreover, their numerical cluster labels are arbitrary. For example, cluster 1 in one draw need not contain the same sites as cluster 1 in another. Predictive probabilities therefore cannot be averaged directly by cluster label. Instead, we distribute each draw-specific cluster's predictive probability among the VI groups according to the proportion of its observed sites belonging to each group.

Let $\mathcal{V}_1,\ldots,\mathcal{V}_{K_{\mathrm{VI}}}$ denote the groups of the VI-optimal partition and $\mathcal{C}^{(b)}_1,\ldots,\mathcal{C}^{(b)}_{K_b}$ the occupied clusters in posterior draw $b$. Define
\[
  q_{hk}^{(b)}
  =
  \frac{|\mathcal{C}^{(b)}_h\cap\mathcal{V}_k|}
       {|\mathcal{C}^{(b)}_h|},
  \qquad k=1,\ldots,K_{\mathrm{VI}}.
\]
Thus, $q_{hk}^{(b)}$ is the proportion of sites in draw-specific cluster $\mathcal{C}^{(b)}_h$ that belong to reference group $\mathcal{V}_k$.

Let $\pi_{0h}^{(b)}$ be the posterior predictive probability that an unmonitored location $\bs{s}_0$ joins occupied cluster $\mathcal{C}^{(b)}_h$, and let $\pi_{0,\mathrm{new}}^{(b)}$ be the probability of forming a new singleton cluster. The predictive probability associated with reference group $\mathcal{V}_k$ is then
\begin{equation*}
\label{eq:supp_vi_projection}
  \widetilde{\pi}_{0k}^{(b)}
  =
  \sum_{h=1}^{K_b}\pi_{0h}^{(b)}q_{hk}^{(b)},
  \qquad k=1,\ldots,K_{\mathrm{VI}}.
\end{equation*}
For example, if a draw-specific cluster contains 80\% of its sites from one VI group and 20\% from another, its predictive probability is divided between those reference groups in the same proportions. This provides a common interpretation across draws without requiring a one-to-one correspondence between their clusters. The new-cluster probability $\pi_{0,\mathrm{new}}^{(b)}$ is retained as a separate category. Averaging over the $B_p$ posterior draws used for these summaries gives
\[
  \bar\pi_{0k}
  =
  \frac{1}{B_p}\sum_{b=1}^{B_p}\widetilde{\pi}_{0k}^{(b)},
  \qquad
  \bar\pi_{0,\mathrm{new}}
  =
  \frac{1}{B_p}\sum_{b=1}^{B_p}\pi_{0,\mathrm{new}}^{(b)}.
\]
These probabilities sum to one and are unchanged by permutations of the draw-specific cluster labels.

In the European analysis, $K_{\mathrm{VI}}=119$, giving 120 categories after including the new-cluster component. Predictive regime certainty is defined as
\begin{equation*}
\label{eq:supp_regime_certainty}
  C(\bs{s}_0)
  =
  \max\left\{
    \bar\pi_{01},\ldots,\bar\pi_{0K_{\mathrm{VI}}},
    \bar\pi_{0,\mathrm{new}}
  \right\}.
\end{equation*}
It is the largest of the averaged predictive probabilities associated with the VI reference groups and the new-cluster category. Higher values indicate stronger concentration on one category, whereas lower values indicate that several alternatives remain plausible. With 120 categories, the uniform-probability benchmark is $1/120\approx0.0083$. A maximum probability of 0.10 therefore indicates concentration relative to this benchmark, but still leaves substantial probability assigned to alternative categories.

A complementary measure is the normalised entropy,
\begin{equation*}
\label{eq:supp_regime_entropy}
  H_{\mathrm{norm}}(\bs{s}_0)
  =
  -\frac{
    \displaystyle\sum_{k=1}^{K_{\mathrm{VI}}}
      \bar\pi_{0k}\log\bar\pi_{0k}
    +
    \bar\pi_{0,\mathrm{new}}\log\bar\pi_{0,\mathrm{new}}
  }{\log(K_{\mathrm{VI}}+1)},
\end{equation*}
with the convention $0\log(0)=0$. It equals zero when all predictive probability is assigned to one category and one when the probabilities are uniform across categories. Certainty describes the support for the leading category, whereas entropy describes how probability is distributed across all alternatives. Neither measure describes uncertainty in PM10 concentration itself, which also involves spatial interpolation, local innovations, regression coefficients, and measurement error.

Predictive means of the temporal parameters can be calculated directly without referring to the VI groups. At each posterior draw, the parameters $(\phi_h^{*(b)},\tau_h^{2*(b)})$ of each occupied cluster are weighted by its predictive allocation probability $\pi_{0h}^{(b)}$, with the new-cluster contribution integrated over the base measures. Averaging these contributions over draws gives probability-weighted summaries of local persistence and innovation variability, rather than the parameters of the dominant reference group alone. These predictive means can therefore vary smoothly over space even though individual regime allocations are discrete.

The European predictive-regime summaries were computed from 1,000 systematically selected posterior draws. To assess the numerical agreement with the original 10,000-draw online summaries, we compared the predictive parameter means obtained by the two calculations. Their correlations were 0.99981 for $\phi$ and 0.99937 for $\tau^2$, with root mean squared discrepancies of 0.00104 and 0.00679, respectively. These comparisons support the numerical agreement of the predictive parameter means; they do not directly assess the Monte Carlo precision of the reference-group probabilities, certainty, or entropy.

\section{Additional simulation summaries}
\label{supp:additional_simulation}

The simulation experiments used fixed prior hyperparameters $(a_\sigma,b_\sigma)=(3,0.16)$, $(a_\xi,b_\xi)=(2,2)$, $(a_\tau,b_\tau)=(2,0.35)$, $(a_\phi,b_\phi)=(1,1)$, $(a_\kappa,b_\kappa)=(3,2)$ and $(a_{\sigma_u},b_{\sigma_u})=(3,0.70)$, with cohesion mass $M=0.5$. These settings were held constant across scenarios and replications. The spatial similarity used standardised coordinates, $\lambda_0=1$, $\nu_0=4$, $\bs{\Lambda}_0=\bs{I}_2$, and $\bs{m}_0$ equal to the empirical mean of the coordinates entering the similarity function.

The main manuscript reports partition recovery, continuous-parameter recovery, predictive RMSE and coverage, and predictive regime classification. Table~\ref{tab:supp_prediction_scores} provides additional summaries of prediction bias, mean absolute error, continuous ranked probability score (CRPS) \citep{gneiting2007strictly}, and predictive interval width. Each entry gives the range of a performance measure across the nine combinations of geographic design and regime separation, with each scenario evaluated over 100 independently fitted datasets. Prediction bias is close to zero for all three processes. Mean absolute error and CRPS quantify predictive performance beyond bias, with lower values indicating better performance. The wider intervals for $w$ and $y$ occur in scenarios with greater local innovation variability, reflecting increased predictive uncertainty rather than systematic over- or underprediction.

\begin{table}[t]
\centering
\footnotesize
\caption{
Range of additional prediction metrics over the nine simulation scenarios.
MAE denotes mean absolute error, CRPS the continuous ranked probability score,
and width the mean width of the 95\% predictive interval.
}
\label{tab:supp_prediction_scores}
\begin{tabular}{lrrrr}
\toprule
Process & Bias & MAE & CRPS & 95\% width\\
\midrule
Common shock $u$ & $[-0.002,0.003]$ & $[0.281,0.288]$ & $[0.199,0.204]$ & $[1.360,1.403]$\\
Latent state $w$ & $[-0.004,0.006]$ & $[0.671,0.996]$ & $[0.475,0.747]$ & $[3.260,6.210]$\\
Response $y$     & $[-0.004,0.005]$ & $[0.708,1.030]$ & $[0.501,0.768]$ & $[3.439,6.272]$\\
\bottomrule
\end{tabular}
\end{table}

\section{Comparative simulation experiments}
\label{supp:comparative_simulation}

\subsection{Competing model specifications}
\label{supp:comparative_simulation_models}

We compared four models throughout. The proposed \emph{DCS-sPPM} is the only
specification that contains all three defining ingredients: a continuously
varying spatial common shock, heterogeneous temporal regimes, and a spatially
informed prior for those regimes. The \emph{PPM with common shock} retains the spatial common-shock
process but removes the geographic similarity term from the partition prior.
The \emph{single-regime model} fixes $K=1$ while retaining the common shock,
and therefore removes heterogeneity in $(\phi_i,\tau_i^2)$. Finally, the
\emph{sPPM without common shock} retains spatially informed temporal regimes
but sets $u_t(\bs{s})\equiv0$. These comparisons isolate the contributions of
spatial cohesion, temporal-regime heterogeneity, and continuously varying
spatial dependence, respectively. Thus, each competitor removes one defining feature of the proposed model, allowing us to assess its contribution to clustering recovery and predictive performance. All models were fitted to the same 100
replicated datasets in each of the nine geography--separation scenarios used
in the main simulation study. Because the data were generated under the full
decomposition, the exercise is a controlled structural-ablation study rather
than a claim that one specification must dominate under every possible data-
generating mechanism.

\subsection{Comparative recovery of temporal regimes}
\label{supp:comparative_clustering}

Table~\ref{tab:supp_comparative_ari} reports mean ARI for the proposed model
and the two comparators that estimate a nontrivial partition. The fourth model,
$K=1$ with common shock, cannot recover a multi-regime partition by
construction and is therefore evaluated for continuous prediction in the next
subsection. When regimes are moderately or clearly separated,
all three models recover the partition accurately. The differences are most
informative in the deliberately difficult, weak-separation setting. For
regular strips and irregular spatial regions, the DCS-sPPM attains mean ARI
values of 0.967 and 0.924, compared with 0.665 and 0.671 under the non-spatial
PPM. Thus, spatial cohesion supplies useful information precisely when the
temporal likelihood alone is weak. Under random allocation, where location is uninformative about the true regime labels, recovery is comparable for the DCS-sPPM and non-spatial PPM (mean ARI 0.654 and 0.687, respectively). The benefit of the spatial partition prior is therefore evident in the geographically coherent settings, but not when the true regimes are unrelated to location.

Removing the common shock is also consequential. Although the no-shock model
performs well for spatially organised regimes, its mean ARI falls to 0.433 in
the random, weak-separation design. In this case, smooth spatial variation that
should be represented by $u_t(\bs{s})$ must instead be accommodated by the
partition and local innovations. The comparison supports the intended
decoupling: the common shock absorbs continuous spatial dependence, leaving
the partition to describe discrete temporal dynamics.

\begin{table}[t]
\centering
\footnotesize
\caption{Mean adjusted Rand index over 100 replications for comparative
partition recovery. The single-regime comparator is omitted because it does
not estimate a nontrivial partition.}
\label{tab:supp_comparative_ari}
\begin{tabular}{llrrr}
\toprule
Geography & Separation & DCS-sPPM & PPM + shock & sPPM, no shock\\
\midrule
Regular strips & Similar      & 0.967 & 0.665 & 0.947\\
Regular strips & Intermediate & 0.999 & 0.999 & 0.997\\
Regular strips & Different    & 1.000 & 0.999 & 0.996\\
Irregular regions & Similar      & 0.924 & 0.671 & 0.870\\
Irregular regions & Intermediate & 1.000 & 0.999 & 0.992\\
Irregular regions & Different    & 1.000 & 1.000 & 0.990\\
Random allocation & Similar      & 0.654 & 0.687 & 0.433\\
Random allocation & Intermediate & 0.991 & 0.999 & 0.996\\
Random allocation & Different    & 0.995 & 0.999 & 0.986\\
\bottomrule
\end{tabular}
\end{table}

The graphical comparison is presented in main-text Figure~\ref{fig:comparative_partition_recovery}; Table~\ref{tab:supp_comparative_ari} provides its numerical values.

\subsection{Comparative prediction}
\label{supp:comparative_prediction}

The prediction comparison evaluates all four models using the same 900
replicated training datasets and held-out locations described in
Section~\ref{sec:simulation}. Table~
\ref{tab:comparative_prediction} in the main manuscript summarises ranges over the nine
scenarios. The proposed DCS-sPPM is the only model that simultaneously
predicts the continuous common shock and a spatially informed local regime.
The DCS-sPPM and PPM with common shock are nearly indistinguishable
for continuous prediction when the temporal regimes are well informed by the
observed trajectories. This is expected: both contain the correctly specified
spatial common shock. The distinction between them concerns prediction of the
discrete regime at a new location, for which only the spatially informed
partition can exploit geographic organisation.

The single-regime model and the model without a common shock illustrate the consequences of omitting temporal heterogeneity and shared spatial variation, respectively. Fixing
$K=1$ worsens common-shock prediction in some scenarios and produces
undercoverage, with response coverage as low as 0.922. Removing the common
shock has a larger and systematic effect on the latent and response surfaces:
the RMSE ranges increase from $[0.847,1.448]$ to $[1.014,1.576]$ for $w$ and from
$[0.892,1.475]$ to $[1.051,1.599]$ for $y$. The resulting intervals remain close to
nominal coverage because they widen to absorb the omitted spatial component;
calibration is therefore achieved at the cost of less precise prediction.
Together with the regime-classification results in the main text, these
experiments show that continuous spatial prediction and predictive regime
allocation are distinct tasks and that the full model provides both within a
single coherent posterior construction.

\section{Exact Gaussian process versus NNGP computation}
\label{sec:nngp_vs_exact}

This section compares the computational cost of the exact Gaussian-process and NNGP spatial calculations, evaluated separately from the full MCMC sampler. We used $m=10$ neighbours for the NNGP representation and performed five evaluations at each network size. Table~\ref{tab:nngp_benchmark} reports the median time per evaluation.

\begin{table}[t]
\centering
\footnotesize
\caption{
Exact Gaussian-process and NNGP timing benchmark.
Times are seconds per evaluation; speedup is the exact-GP time divided by
the NNGP time.
}
\label{tab:nngp_benchmark}
\begin{tabular}{rrrr}
\toprule
$n$ & Exact GP & NNGP ($m=10$) & Speedup\\
\midrule
100  & 0.0022 & 0.0041 & 0.54\\
500  & 0.0593 & 0.0191 & 3.11\\
1000 & 0.2890 & 0.0375 & 7.71\\
2000 & 1.8390 & 0.0784 & 23.46\\
3000 & 5.1630 & 0.1187 & 43.51\\
4000 & 11.3030 & 0.1462 & 77.30\\
\bottomrule
\end{tabular}
\end{table}

At $n=100$, dense Gaussian-process computation remains slightly faster
because sparse-structure setup dominates the small problem. The NNGP
representation becomes faster by $n=500$ and its advantage increases rapidly
with network size. At $n=4{,}000$, the retained benchmark is approximately
77 times faster under the NNGP representation. The estimated storage for the dense covariance matrix and its Cholesky factor is approximately 244 MiB, compared with 0.49 MiB for the stored NNGP coefficients, conditional variances and neighbour indices. These figures describe the stored representations rather than peak memory usage.

These results support the use of the nearest-neighbour representation in the
continental PM10 analysis, where the observed network contains 2,989 monitoring locations.

\section{Additional details for the European PM10 analysis}
\label{supp:pm10}

\subsection{Data construction and additional exploratory summaries}
\label{supp:pm10_data}

The analysis uses daily PM10 measurements during the 2024 calendar year. The study domain was restricted to the longitude--latitude bounding box $[-11,46]\times[34,72]$, covering European countries together with Turkey and Cyprus, but excluding Iceland. PM10 observations and station metadata were obtained through the European Environment Agency (EEA) Air Quality download. PM10 observations were downloaded as native daily and hourly aggregation products. Records with \texttt{Validity=1} and positive concentrations were retained; native daily records were additionally restricted to concentrations below $500\,\mu\mathrm{g}/\mathrm{m}^3$. Hourly records were averaged by day. When native daily and hourly-derived values were both available for the same sampling point and day, the lower value was retained. Monitoring coordinates and station descriptors were matched from the EEA metadata using sampling-point identifiers.

Series were eligible when observations were available for at least 90\% of the 366 days. Remaining missing days were filled by linear interpolation, with endpoint gaps extended using the nearest available observation. Where multiple series represented the same monitoring station or shared identical coordinates, one eligible series was retained according to a predefined ranking based on measurement method and documented equivalence to reference methods, followed by data provenance (native daily or hourly-derived measurements), temporal coverage and the length of the longest missing interval. Ambiguous cases were reviewed individually. The final analysis comprises 2,989 series at unique monitoring locations, used consistently for all four models. PM10 concentrations are expressed in $\mu\mathrm{g}/\mathrm{m}^3$.

The meteorological covariates were obtained from preprocessed ERA5 reanalysis fields on a regular $0.25^\circ\times0.25^\circ$ longitude--latitude grid for all 366 days of 2024. Temperature is the daily-mean 2-m air temperature in degrees Celsius, and precipitation is the retrieved daily-mean total-precipitation field in metres. The wind covariate is the magnitude of the vector formed by the daily-mean zonal and meridional 10-m wind components, expressed in metres per second. Each monitoring series was matched to its nearest available ERA5 grid cell. Before fitting, PM10 and the three meteorological covariates were centred and scaled using fixed empirical constants stored in the analysis preparation metadata. The same transformations were used across all four models and for their prediction inputs. Response predictions were subsequently transformed back to the original concentration units.

A $100\times100$ longitude--latitude grid was intersected with the medium-resolution Natural Earth country polygons in WGS84 (EPSG:4326), retaining 5,297 land-based prediction locations within the study domain. The sPPM similarity used spatial coordinates centred and divided by their empirical standard deviations, whereas NNGP distances were calculated after an approximate metric transformation of longitude and latitude. At each prediction location, meteorological covariates were taken from the ERA5 cell associated with the nearest retained monitoring station. 

\begin{figure}[t]
    \centering
    \includegraphics[width=\textwidth]
    {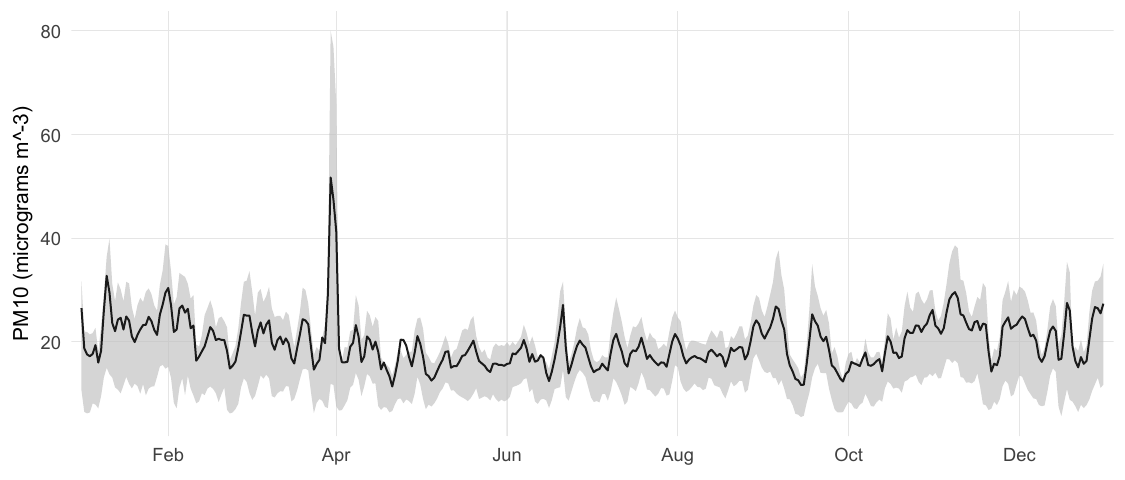}
    \caption{
    Daily empirical summary of PM10 concentrations across the European
    monitoring network during 2024. The solid line shows the daily mean
    across retained monitoring series and the shaded band represents the
    corresponding cross-station interquartile range. 
    }
    \label{fig:europe_temporal_summary}
\end{figure}

Figure~\ref{fig:europe_temporal_summary} complements the spatial exploratory summaries in the main text by showing the daily mean and interquartile range of PM10 across the retained monitoring network.

\subsection{Additional response prediction maps}
\label{supp:response_prediction_maps}

\begin{figure}[t]
\centering
\begin{overpic}[width=\textwidth]{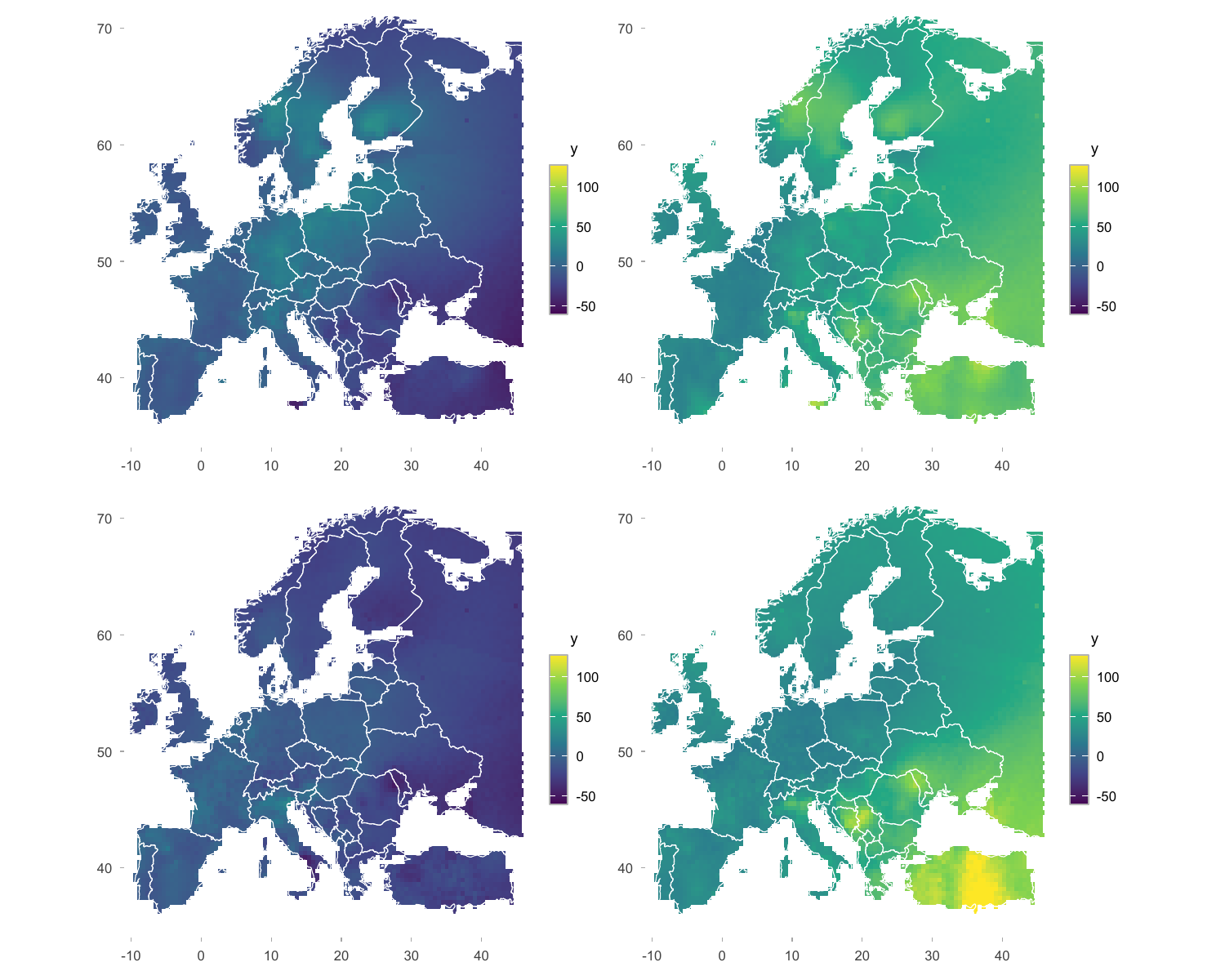}
\put(1.4,77.9){\jrsscpanellabel{A}}
\put(51.4,77.9){\jrsscpanellabel{B}}
\put(1.4,37.9){\jrsscpanellabel{C}}
\put(51.4,37.9){\jrsscpanellabel{D}}
\end{overpic}
\caption{Gaussian-derived 95\% predictive limits for PM10: lower 2.5\% limits (A, C) and upper 97.5\% limits (B, D) for 9 April (A, B) and 15 December (C, D) 2024. All quantities are expressed in $\mu\mathrm{g}/\mathrm{m}^3$, using a common concentration colour scale across all four panels.}
\label{fig:response_predictive_limits}
\end{figure}

Figure~\ref{fig:response_predictive_limits} presents the lower and upper limits corresponding to the response means and standard deviations reported in the main manuscript for April 9 and December 15, 2024. The limits are constructed as the predictive mean plus or minus $1.96$ times the predictive standard deviation. They are pointwise, moment-matched Gaussian approximations, rather than empirical posterior quantiles or simultaneous uncertainty bands; negative lower limits can therefore occur even though PM10 concentration is non-negative.

\subsection{Application computation and diagnostics}
\label{supp:application_diagnostics}

The European analysis used $m=10$ NNGP neighbours and max--min ordering.
The inverse-Gamma hyperparameters were
$(a_\sigma,b_\sigma)=(3,0.40)$,
$(a_\xi,b_\xi)=(2,2)$,
$(a_\tau,b_\tau)=(2,0.35)$, and
$(a_{\sigma_u},b_{\sigma_u})=(3,0.60)$.
We used
$(a_\phi,b_\phi)=(1,1)$,
the shape--rate Gamma prior
$(a_\kappa,b_\kappa)=(3,2)$,
and DP cohesion mass $M=1$. For the spatial similarity, we used $\lambda_0=1$, $\nu_0=4$, $\bs{\Lambda}_0=\bs{I}_2$, and $\bs{m}_0$ equal to the empirical mean of the standardised monitoring coordinates.

The sampler ran for 30,000 iterations, of which 20,000 were discarded as
burn-in and 10,000 retained, using eight OpenMP threads. Posterior predictive
moments for $u_t(\bs{s})$, $w_t(\bs{s})$, and $y_t(\bs{s})$ were accumulated
online at all 5,297 prediction locations and 366 time points. The DCS-sPPM fit required 13.60 hours, including online prediction but excluding subsequent VI partition summarisation and figure generation, on an Intel(R) Core(TM) i7-10750H CPU @ 2.600GHz machine.

The overall acceptance rates were 0.440 for persistence updates and 0.226 for the spatial move. Effective sample sizes for selected quantities from the 10,000 retained iterations are reported in Table~\ref{tab:supp_application_diagnostics}.

\begin{table}[t]
\centering
\footnotesize
\caption{Effective sample sizes for selected quantities from the 10,000 retained iterations of the unique-location European PM10 analysis.}
\label{tab:supp_application_diagnostics}
\begin{tabular}{lr}
\toprule
Quantity & ESS\\
\midrule
Intercept & 33.6\\
Temperature & 53.9\\
Precipitation & 100.6\\
Wind speed & 74.0\\
$\sigma^2$ & 362.7\\
$\sigma_u^2$ & 1,569.4\\
$\kappa$ & 1,258.1\\
Number of clusters & 88.1\\
Log-likelihood & 353.0\\
\bottomrule
\end{tabular}
\end{table}

Effective sample sizes vary across inferential quantities, with larger values for the spatial covariance parameters than for the regression coefficients and cluster count. The table reports these differences explicitly; the diagnostics describe the retained single chain.

Table~\ref{tab:supp_application_timing} reports the computational profile of the final analysis. Temporal-regime updates accounted for approximately 48\% of runtime. Online accumulation and diagnostics accounted for a further 17.8\%, allowing predictive summaries to be obtained without retaining individual predictive draws over all locations and days.

\begin{table}[t]
\centering
\footnotesize
\caption{Block-specific timing for the European analysis. Times are expressed in hours.}
\label{tab:supp_application_timing}
\begin{tabular}{lrr}
\toprule
Block & Time (h) & Percentage\\
\midrule
Common shocks $u$                    & 0.55 & 4.1\\
Regression coefficients              & 0.26 & 1.9\\
Latent state and measurement variance & 0.38 & 2.8\\
Allocation update                    & 1.82 & 13.4\\
Temporal-regime update                & 6.53 & 48.0\\
Spatial parameters                   & 1.56 & 11.5\\
Online storage and diagnostics       & 2.42 & 17.8\\
Other                                & 0.07 & 0.5\\
\bottomrule
\end{tabular}
\end{table}

\section{Comparative European PM10 analyses}
\label{supp:comparative_application}

\subsection{Design and computation}
\label{supp:comparative_application_design}

We fitted all four models---the proposed DCS-sPPM and the three structural comparators used in the simulation study---to the same European dataset. The response, covariates, monitoring network, prediction grid and nominal MCMC settings were held fixed, with common prior specifications for shared parameters and the same NNGP configuration where applicable. Each analysis used 30,000 iterations, discarded the first 20,000, and accumulated online predictions over all 5,297 land-grid locations and 366 days during the final 10,000 iterations. Equal iteration counts do not imply equal effective sample sizes, however, so model-specific Monte Carlo limitations remain relevant to interpretation.

Elapsed times were 13.60 hours for DCS-sPPM, 10.82 hours for PPM with a common shock, 4.76 hours for the single-regime model with a common shock, and 9.88 hours for sPPM without a common shock. These timings include online prediction but exclude subsequent VI partition summarisation and figure generation. Overall persistence-update acceptance ranged from 0.425 to 0.441; spatial acceptance ranged from 0.211 to 0.226 among models containing a common shock. Effective sample sizes for $\kappa$ and $\sigma_u^2$ were 1,258 and 1,569 for DCS-sPPM, 1,168 and 1,324 for PPM with a common shock, and 650 and 917 for the single-regime model.

\subsection{Partition structure and variance allocation}
\label{supp:comparative_application_structure}

Table~\ref{tab:supp_application_structure} compares the four models fitted to the same dataset. The DCS-sPPM yields a posterior mean of 107.6 occupied groups and a VI representative partition with 119 groups. Removing geographic similarity reduces the posterior mean cluster count to 93.4 and produces more spatially dispersed representative groups: median within-cluster distance increases from 644 to 733 km. Without the common shock, the median distance is 401 km. The tighter grouping under this restriction is consistent with a redistribution of spatial variation into the local temporal structure, rather than demonstrating superior temporal-regime identification. The full model retains separate representations of shared spatial variation and heterogeneous local dynamics.

\begin{table}[t]
\centering
\footnotesize
\caption{Posterior partition summaries in the comparative European analyses.
Distances refer to the representative VI partition.}
\label{tab:supp_application_structure}
\begin{tabular}{lrrrrr}
\toprule
Model & Mean $K$ & 95\% interval & VI $K$ & Singletons & Median distance (km)\\
\midrule
DCS-sPPM       & 107.6 & [102,113] & 119 & 20 & 644\\
PPM + shock    & 93.4  & [89,99]   & 92  & 15 & 733\\
$K=1$ + shock  & 1.0   & [1,1]     & 1   & 0  & ---\\
sPPM, no shock & 101.1 & [97,105]  & 112 & 18 & 401\\
\bottomrule
\end{tabular}
\end{table}

The DCS-sPPM and PPM with a common shock yield similar common-shock variances, whereas imposing a single temporal regime increases both the measurement-error and common-shock variances. This indicates that the estimated allocation of variability among model components depends on whether heterogeneous local temporal dynamics are represented.

\subsection{Consequences for spatial prediction}
\label{supp:comparative_application_prediction}

Because responses are observed only at monitoring locations, the comparisons at unmeasured locations are structural sensitivity analyses rather than out-of-sample
predictive scores. We report all four models, taking the proposed DCS-sPPM as
the reference because it contains the complete decomposition. Table~
\ref{tab:supp_application_surfaces} reports correlations and RMSEs on the
standardised response scale.

Main-text Figure~\ref{fig:comparative_application} summarises these comparisons: panel (A) shows the agreement between each competitor's posterior mean prediction surfaces and those of DCS-sPPM, while panel (B) compares the estimated variance components relative to the full model.

Removing geographic similarity from the partition prior leaves the common-shock posterior mean surface almost unchanged, with correlation 0.999 relative to DCS-sPPM; the corresponding correlations for the latent state and response are 0.986 and 0.987. Predictive uncertainty nevertheless differs between the models. Comparing their posterior predictive standard deviations at each location and time gives RMSE values of 0.480 for $w$ and 0.479 for $y$ on the standardised scale. The PPM also yields approximately 9\% higher average predictive standard deviations for these processes. Thus, similar predicted means can be accompanied by different assessments of uncertainty.

Imposing a single temporal regime or removing the common shock produces larger differences in the predicted processes relative to DCS-sPPM. Under a single temporal regime, response-surface correlation is 0.874 and average common-shock predictive standard deviation is approximately 30\% larger. Without the common shock, latent-state mean-surface correlation is close to zero, response correlation is 0.290, and average predictive standard deviations for $w$ and $y$ are approximately 20\% larger. Together with the simulation results, these comparisons support retaining both continuous spatial variation and heterogeneous temporal dynamics, while showing that spatial cohesion principally changes regime organisation and local predictive uncertainty. 

\begin{table}[t]
\centering
\footnotesize
\caption{Posterior predictive surface comparison for all four models, using
the proposed DCS-sPPM as the full-model reference. Correlation and RMSE refer
to posterior means; SD RMSE compares posterior standard-deviation surfaces.}
\label{tab:supp_application_surfaces}
\begin{tabular}{llrrr}
\toprule
Model & Process & Mean correlation & Mean RMSE & SD RMSE\\
\midrule
DCS-sPPM       & $u$ & 1.000 & 0.000 & 0.000\\
               & $w$ & 1.000 & 0.000 & 0.000\\
               & $y$ & 1.000 & 0.000 & 0.000\\
PPM + shock    & $u$ & 0.999 & 0.016 & 0.004\\
               & $w$ & 0.986 & 0.106 & 0.480\\
               & $y$ & 0.987 & 0.104 & 0.479\\
$K=1$ + shock  & $u$ & 0.857 & 0.211 & 0.099\\
               & $w$ & 0.873 & 0.448 & 0.396\\
               & $y$ & 0.874 & 0.419 & 0.392\\
sPPM, no shock & $w$ & $-0.005$ & 0.585 & 0.346\\
               & $y$ & 0.290 & 0.603 & 0.345\\
\bottomrule
\end{tabular}
\end{table}

Main-text Figure~\ref{fig:comparative_application} summarises the surface agreement and relative variance allocation across the four specifications. The tables above provide the detailed numerical comparisons.

\section{Additional predictive temporal-regime summaries}
\label{supp:pm10_predictive_regimes}

\begin{figure}[t]
\centering
\begin{overpic}[width=\textwidth]{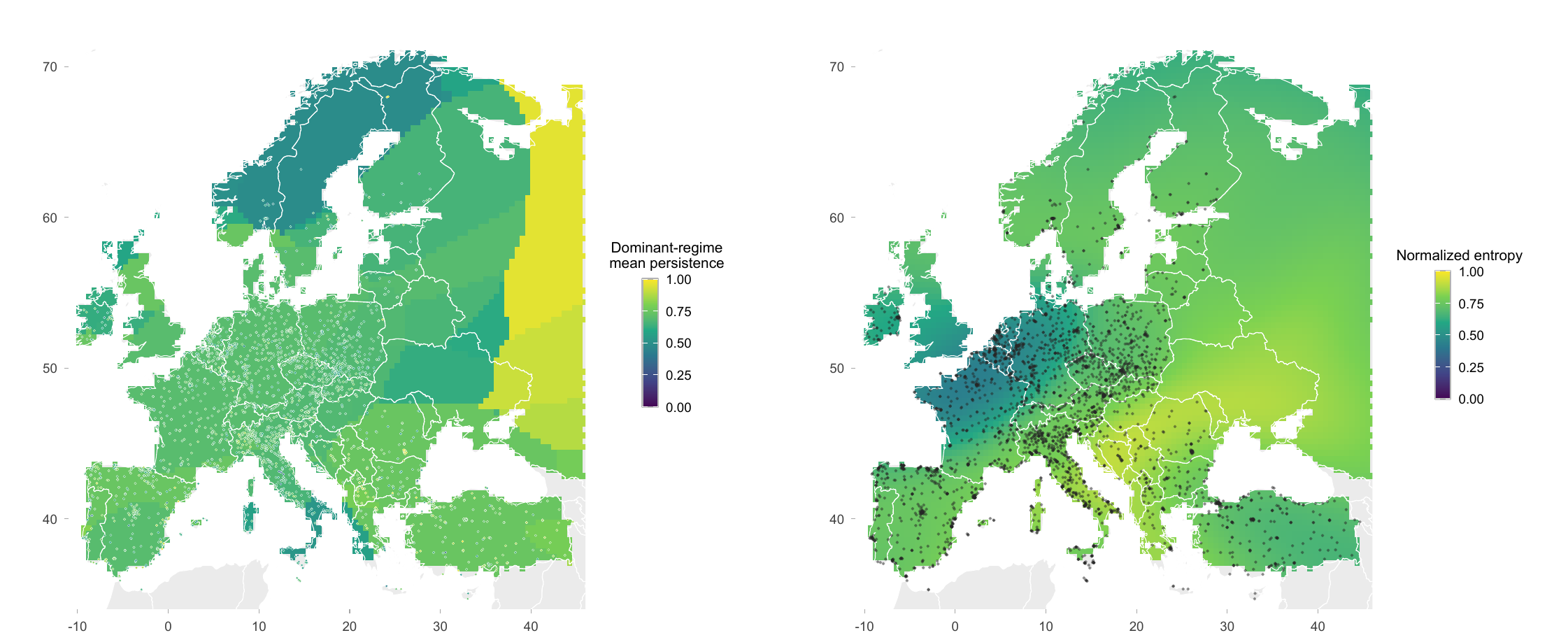}
\put(0.75,39.3){\jrsscpanellabel{A}}
\put(51.3,39.3){\jrsscpanellabel{B}}
\end{overpic}
\caption{Additional predictive regime summaries: (A) the dominant VI-projected regime, coloured by its representative persistence; (B) normalised predictive regime entropy. The dominant-regime map should be interpreted alongside allocation certainty rather than as a sharply determined segmentation.}
\label{fig:predictive_regime_map}
\end{figure}

Figure~\ref{fig:predictive_regime_map} provides the categorical counterpart
to the label-invariant predictive parameter surfaces shown in the main text.
At each prediction location, the dominant VI regime is the representative
group receiving the largest projected predictive probability. Colours are
indexed by the representative persistence parameter to preserve some ordering
among the large number of categories. The normalised entropy map provides a
complementary view of allocation uncertainty to the maximum-probability
certainty surface shown in the main text.

Across prediction locations, the largest projected regime probability has median 0.119 and interquartile range $[0.090,0.163]$, while normalised entropy has median 0.735 and interquartile range $[0.684,0.788]$. These summaries indicate that several temporal regimes generally remain plausible at an unmonitored location. Accordingly, the dominant-regime map represents the most supported category rather than a confident classification, whereas the predictive parameter surfaces average over the competing regime assignments.

\bibliographystyle{apalike}
\bibliography{reference}

\end{document}